\documentclass[final,3p]{elsarticle}

\usepackage{amssymb}
\usepackage [table]{xcolor}
\usepackage{setspace}
\usepackage{pdfpages}
\usepackage{amsmath}
\usepackage{color,soul}
\usepackage{multirow}
\usepackage{graphicx}
\usepackage{parskip}
\usepackage{caption}
\usepackage{subcaption}
\usepackage{textcomp}

\makeatletter
\def\ps@pprintTitle{%
  \let\@oddhead\@empty
  \let\@evenhead\@empty
  \let\@oddfoot\@empty
  \let\@evenfoot\@oddfoot
}
\makeatother

\journal{Comm Biology}
\begin{document}
\begin{frontmatter}

\title{On the characteristics of natural hydraulic dampers: An image-based approach to study the fluid flow behaviour inside the human meniscal tissue.}

\author[a]{Waghorne J.}
\author[b]{Bonomo F.P.}
\author[c]{Rabbani A.}
\author[a]{Bell D.}
\author[a,d]{Barrera O.}

\address[a]{School of Engineering, Computing and Mathematics, Oxford Brookes University, Oxford, United Kingdom}
\address[b]{Advanced Technology Network Center (ATeN Center), Università degli Studi di Palermo, 90128 Palermo, Italy}
\address[c]{School of Computing, University of Leeds, Leeds, UK}
\address[d]{Department of Engineering Science, University of Oxford}


\cortext[cor1]{To whom correspondence should be addressed. E-mail: olga.barrera@eng.ox.ac.uk, obarrera@brookes.ac.uk}



\begin{abstract}

The meniscal tissue is a layered material with varying properties influenced by collagen content and arrangement. Understanding the relationship between structure and properties is crucial for disease management, treatment development, and biomaterial design. The internal layer of the meniscus is softer and more deformable than the outer layers, thanks to interconnected collagen channels that guide fluid flow. To investigate these relationships, we propose a novel approach that combines Computational Fluid Dynamics (CFD) with Image Analysis (CFD-IA). We analyze fluid flow in the internal architecture of the human meniscus across a range of inlet velocities (0.1mm/s to 1.6m/s) using high-resolution 3D micro-computed tomography scans. Statistical correlations are observed between architectural parameters (tortuosity, connectivity, porosity, pore size) and fluid flow parameters (Re number distribution, permeability). Some channels exhibit Re values of 1400 at an inlet velocity of 1.6m/s, and a transition from Darcy's regime to a non-Darcian regime occurs around an inlet velocity of 0.02m/s. Location-dependent permeability ranges from 20-32 Darcy. Regression modelling reveals a strong correlation between fluid velocity and tortuosity at high inlet velocities, as well as with channel diameter at low inlet velocities. At higher inlet velocities, flow paths deviate more from the preferential direction, resulting in a decrease in the concentration parameter by an average of 0.4. This research provides valuable insights into the fluid flow behaviour within the meniscus and its structural influences.

\end{abstract}
\end{frontmatter}

\section{Significance}

The knee meniscus is a highly porous soft tissue with remarkable properties of load transfer from the upper to lower part of the body. The structure is similar to a sandwich structure with a stiff outside layer and a soft internal layer, enabling it to accommodate deformation and dissipate energy, making it a potentially optimised damping system. The secret of the internal layer is the arrangement of the collagen in a network of channels oriented in a preferential direction guiding the fluid flow paths. We give insight into the relation between architectural and fluid flow parameters, in addition to fluid regimes, which is essential to design suitable biomimetic systems that can be adopted for replacement. 
 
\section{Introduction}

Nature has optimized the architecture of tissues to fulfil specific functions. For example, the nature-designed lightweight structural materials with optimal strength-to-weight and stiffness-to-weight performance \citep{libonati20213d}. Nature offers several effective solutions, particularly within porous structures, from shock-absorbing hedgehog spines \citep{swift2016hedgemon} to the flexible-protective exoskeleton of the Polypterus Senegalus \citep{zolotovsky2021fish}, to the adaptive adhesion properties of a Gecko's feet \citep{shah2004modeling} and trabecular bone \citep{libonati20213d}; to name a few of the innumerable adaptations found in nature. 
The structure of hedgehog spines is similar to a foam that fills the central part of a spine, and supports the thin outer wall, contrasting local instability and allowing the whole system to bend further without breaking. This type of sandwich structure allows the hedgehog to bounce when it falls from a height, thus preventing injuries without overloading the animal. Trabecular bone tissue is another classic example of cellular structure at the microscale. It has an open-cell porous arrangement, allegedly random, but carefully designed by nature to bear specific local loadings \citep{huiskes1996microdamage} at low weights.
Whilst the structure-function relationship is well-studied in natural hard materials, soft materials are not receiving similar attention, despite their high prevalence in nature \citep{sharabi2022structural}.

\par

Advanced imaging techniques, mechanical testing and models \citep{maritz2020development,agustoni2021high, maritz2021functionally, vetri2019advanced,bulle2021human, gunda2023fractional,sancataldo2022two} have shown that the meniscus, one of the major soft tissues present in the knee joint, presents a through-thickness functionally graded structure consisting of three critical layers: top, middle, bottom. The top and bottom are thin layers that are stiffer and less permeable, the middle layer is thicker, more permeable and softer - to accommodate deformation and dissipate energy - making it an optimised damping system. The way nature has conceived the architecture of the internal layer within the meniscus relies on a hierarchical anisotropic and inhomogeneous network of collagen channels through which fluid flows. Fluid flowing inside these channels determines the time-dependent behaviour of the tissue. As the tissue deforms under the action of loads these channels change morphology resulting in permeability changes as a function of channel dimension, porosity and tortuosity, making the permeability tensor not only dependent on pressure but also varying spatially and temporally because of the local variation of applied pressure gradients. 

\par

There are presently macroscopic models being developed to consider linear and nonlinear changes of permeability due to deformations, for example, the biphasic or poroelastic models for brain tissues \citep{yuan2022microstructurally}. One of the components of poroelastic models is Darcy's law which is used to derive the pore pressure diffusion equation that is needed to be coupled with the stress tensor to model the coupling effect of solid deformation and pore pressure field (which summarises the role of fluid flow in the deformation process). Various researchers proposed possible reasons for this deviation from Darcy's law, i.e. macro-roughness of the pores, viscous boundary layer, kinetic energy losses, microscopic inertial forces manifested in the interstitial drag force, the singularity of streamline patterns and/or the non-periodicity of the micro-scale flow, the variation of integral viscous dissipation due to a deformation of
streamline patterns, and the inhomogeneity of the porous medium which causes higher local fluid velocity. Due to these factors, several flow regimes dependent on local fluid velocity were observed during the percolation of liquid through the porous space.

\par

Recently, a multi-scale model was proposed for Convection-Enhanced Delivery (CED) \citep{yuan2022microstructurally} which takes into account the pressure-driven deformation of brain micro-structure to quantify the change of local permeability and porosity. Results show that both hydraulic pressure and drug concentration in the brain would be significantly underestimated by classical Darcy’s law. Understanding and appropriately modelling the fluid flow behaviour in the different portions of the meniscal tissue is essential to gain insight into the biomechanical function of the tissue.

Image-based permeability calculation methods have the advantage of being able to capture the complex internal geometry of porous media and have the potential for more accurate predictions of fluid flow \citep{rabbani2021review}. There are several common methods for modelling fluid flow in porous media, including solving Navier-Stokes equations \citep{liu2019interior}, lattice Boltzmann method \citep{cheng2019effect,eshghinejadfard2016calculation}, pore network modelling \citep{baychev2019reliability}, solving Laplace equation \citep{da2021fast}, or a combination of the above mentioned \citep{rabbani2019hybrid}. Navier-Stokes equations are a set of partial differential equations that describe the motion of fluids, taking into account fluid viscosity, density, and pressure. This method is often used as a benchmark although it is sensitive to the meshing quality and grid simplifications that are inevitable for intricate geometries \citep{ali2019efficient}. The lattice Boltzmann method is a relatively new and promising approach for modelling fluid flow in porous media. The lattice Boltzmann method simulates the behaviour of fluids by modelling the interaction between particles on a lattice and can use the same discretized space of the image with no need for mesh simplifications \citep{cheng2019effect}. Pore network modelling (PNM) is another method for modelling fluid flow in porous media. PNM represents the porous media as a network of simplified interconnected pores and throats. The flow through the network is modelled by solving equations for the pressure and flow rate in each pore and throat \citep{blunt2001flow}. PNM is computationally efficient, however, due to over-simplification its predictions are less accurate \citep{rabbani2019hybrid}. One major gap in the literature on fluid flow in porous media is the fact that most of the methods originated in geo-sciences and have been developed and validated for tight mediums with low porosity, most of which are below 30\%, which is common in geological porous materials \citep{rabbani2021review}. In highly porous geometries such as miniscal tissue, complex flow patterns are more likely to be observed which needs to be addressed.       

\begin{figure}[h!]
    \centering
    \includegraphics[scale=1.5]{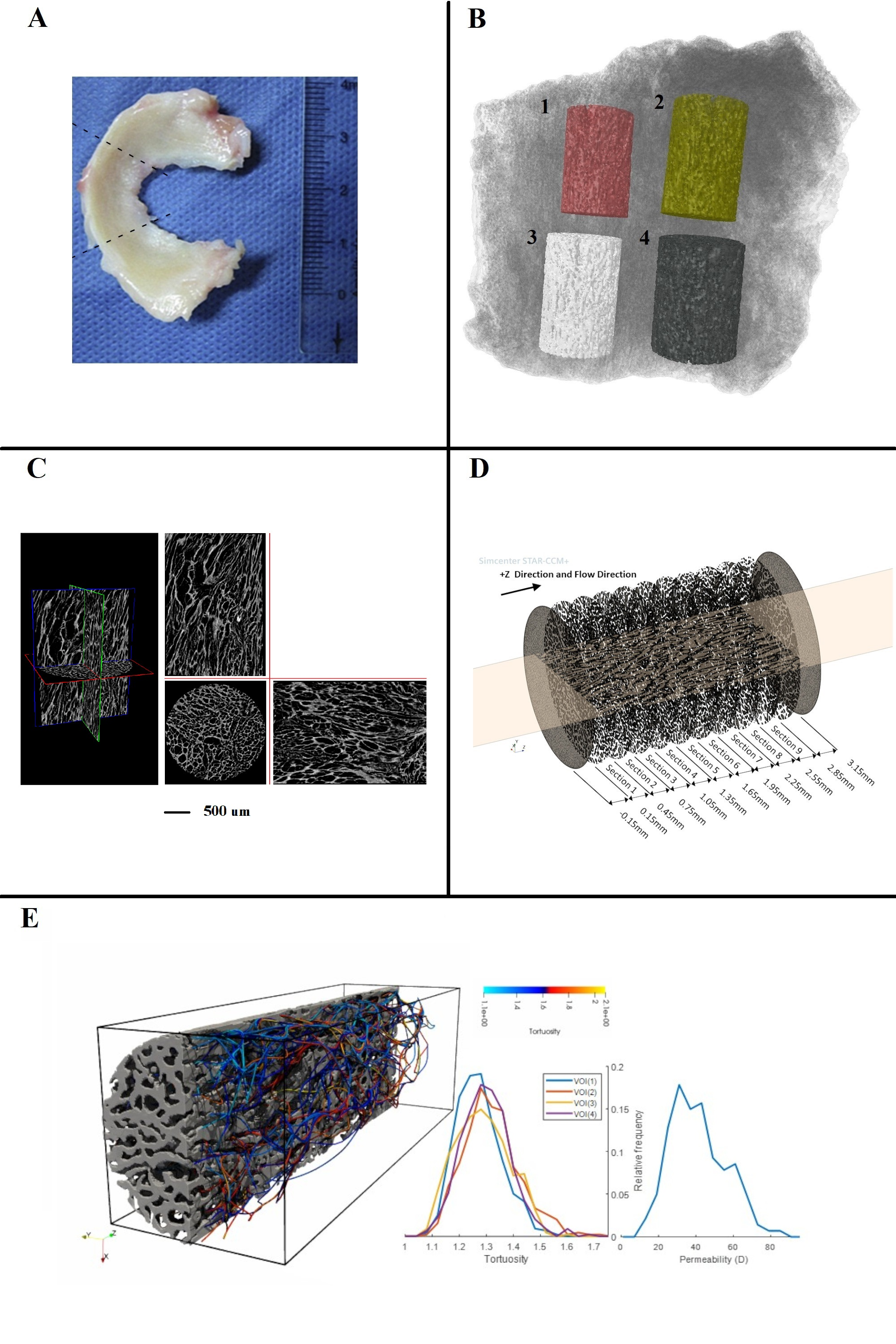}      
    \caption{A. Human lateral meniscus, dashed lines represent the body region of the meniscus object of the present study. B. Regions of interest extracted from the 3D reconstruction of the scan with the following dimensions : 1 = VOI(1)  $2mm\times 3mm$; 2 = VOI(2) $2mm\times 3mm$; 3= VOI(3)  $1mm\times 3.13mm$; 4= VOI(4)  $2.32mm\times 3mm$. C. Sagittal and coronal planes of VOI(4), resolution of $6.25 \mu m$. D. Location of sections long z-direction inside VOI(4). E. Graphical distribution of tortuosity inside VOI(3), distributions of tortuosity in the four datasets, distribution of permeability in VOI(4).}
    \label{fig:PorosityvsL}
\end{figure}
    
In this present study, we present micro-computed image-based pore-scale simulations of flow in native micro-structures which allow visualization of the flow behaviour in digital images. Running numerical Darcy's experiments using a range of inlet velocities from 0.0001m/s to 1.6m/s, allowing for the study of the fluid flow regime from Darcian to the onset of inertial effect. Quantitative pore-scale information such as velocity, local Reynold's number and the link to the morphology of the channels lay the foundation of fluid transport phenomena in the meniscal tissue.

\section{Results} \label{sect: results}

\subsection{Statistical characterization of architectural parameters}
Figure \ref{fig:PorosityvsL} presents the four datasets, VOI - VOI(4), extracted from the body region of the meniscus along the preferential direction of channels. A video demonstration exploring the morphology of the meniscal channels of VOI(4) has been included in the supplementary material. The distributions of tortuosity for the four datasets are reported in Fig.\ref{fig:PorosityvsL}d. It was noted that these distributions are highly consistent throughout the four data sets. This is a feature that reoccurred frequently for many of the architectural parameters, also being true for distributions of porosity, channel diameter and connectivity in the four data sets, as present in the supporting material Fig. S2. Given the similarity in the architecture of the four datasets, we have modelled the fluid flow behaviour in just two datasets: VOI(3) and VOI(4). These two samples were selected as they are the smallest and largest respectively of the data sets, which allows for an investigation into the effect of sample size on results. Furthermore, to relate the evolution of fluid behaviour to microstructural changes inside the domain, linearly spaced subsections of VOI(3) and VOI(4) have been analysed during the post-processing of the simulations. Each of the subsections are consistently $0.4mm$ and $0.3mm$ in length along the flow direction for the VOI(3) and VOI(4) samples respectively. 
\par
Table \ref{table:archi} presents the mean averaged values and standard deviation (SD) of the morphological and topological parameters of the simulated VOI (3) and VOI (4) which have gone through a clean-up process involving the removal of small islands and non-connected fluid space: porosity, channel diameter, connectivity and tortuosity, details of how these values were obtained can be found in Section \ref{sect: method, characterisation}. From Table \ref{table:archi} it is noted that VOI(4) is over 20\% more porous than VOI(3). The variation of this base parameter can also help explain the difference seen in other parameters. For example, pores in VOI(4) on average are connected to three more pores than those in VOI(3), this can be partially explained by the increased porosity, as there is more space for pores to be located and pores generally have higher surface area to connect with other pores. Table \ref{table:archi} also demonstrates that the VOI(4) sample is notably less tortuous than VOI(3). It is also apparent that there is a sudden drop in porosity in the VOI(3) sample from z=2.6 to 3.0mm (towards the end of the sample), due to a change in the tissue fibre structure.

\begin{table}[]
\centering
\resizebox{\textwidth}{!}{%
\begin{tabular}{|c|ccc|ccccccccc|}
\hline
\rowcolor[HTML]{9B9B9B} 
\cellcolor[HTML]{9B9B9B} &
  \multicolumn{3}{c|}{\cellcolor[HTML]{9B9B9B}\textbf{Sample}} &
  \multicolumn{9}{c|}{\cellcolor[HTML]{9B9B9B}\textbf{Section}} \\ \cline{2-13} 
\rowcolor[HTML]{9B9B9B} 
\multirow{-2}{*}{\cellcolor[HTML]{9B9B9B}\textbf{Parameter}} &
  \multicolumn{1}{c|}{\cellcolor[HTML]{9B9B9B}\textbf{Name}} &
  \multicolumn{1}{c|}{\cellcolor[HTML]{9B9B9B}\textbf{Average}} &
 \multicolumn{1}{c|}{\cellcolor[HTML]{9B9B9B}\textbf{SD}} &
  \multicolumn{1}{c|}{\cellcolor[HTML]{9B9B9B}\textbf{1}} &
  \multicolumn{1}{c|}{\cellcolor[HTML]{9B9B9B}\textbf{2}} &
  \multicolumn{1}{c|}{\cellcolor[HTML]{9B9B9B}\textbf{3}} &
  \multicolumn{1}{c|}{\cellcolor[HTML]{9B9B9B}\textbf{4}} &
  \multicolumn{1}{c|}{\cellcolor[HTML]{9B9B9B}\textbf{5}} &
  \multicolumn{1}{c|}{\cellcolor[HTML]{9B9B9B}\textbf{6}} &
  \multicolumn{1}{c|}{\cellcolor[HTML]{9B9B9B}\textbf{7}} &
  \multicolumn{1}{c|}{\cellcolor[HTML]{9B9B9B}\textbf{8}} &
  \multicolumn{1}{c|}{\cellcolor[HTML]{9B9B9B}\textbf{9}} \\ \hline
\rowcolor[HTML]{EFEFEF} 
\cellcolor[HTML]{EFEFEF} &
  \multicolumn{1}{c|}{\cellcolor[HTML]{EFEFEF}VOI(3)} &
  \multicolumn{1}{c|}{\cellcolor[HTML]{EFEFEF}45} &
  5.0 &
  \multicolumn{1}{c|}{\cellcolor[HTML]{EFEFEF}44} &
  \multicolumn{1}{c|}{\cellcolor[HTML]{EFEFEF}47} &
  \multicolumn{1}{c|}{\cellcolor[HTML]{EFEFEF}48} &
  \multicolumn{1}{c|}{\cellcolor[HTML]{EFEFEF}50} &
  \multicolumn{1}{c|}{\cellcolor[HTML]{EFEFEF}50} &
  \multicolumn{1}{c|}{\cellcolor[HTML]{EFEFEF}45} &
  \multicolumn{1}{c|}{\cellcolor[HTML]{EFEFEF}38} &
  \multicolumn{1}{c|}{\cellcolor[HTML]{EFEFEF}-} &
  - \\ \cline{2-13} 
\rowcolor[HTML]{EFEFEF} 
\multirow{-2}{*}{\cellcolor[HTML]{EFEFEF}Porosity (\%)} &
  \multicolumn{1}{c|}{\cellcolor[HTML]{EFEFEF}VOI(4)} &
  \multicolumn{1}{c|}{\cellcolor[HTML]{EFEFEF}66} &
  2.6 &
  \multicolumn{1}{c|}{\cellcolor[HTML]{EFEFEF}61} &
  \multicolumn{1}{c|}{\cellcolor[HTML]{EFEFEF}64} &
  \multicolumn{1}{c|}{\cellcolor[HTML]{EFEFEF}67} &
  \multicolumn{1}{c|}{\cellcolor[HTML]{EFEFEF}68} &
  \multicolumn{1}{c|}{\cellcolor[HTML]{EFEFEF}68} &
  \multicolumn{1}{c|}{\cellcolor[HTML]{EFEFEF}68} &
  \multicolumn{1}{c|}{\cellcolor[HTML]{EFEFEF}68} &
  \multicolumn{1}{c|}{\cellcolor[HTML]{EFEFEF}67} &
  66 \\ \hline
\rowcolor[HTML]{C0C0C0} 
\cellcolor[HTML]{C0C0C0} &
  \multicolumn{1}{c|}{\cellcolor[HTML]{C0C0C0}VOI(3)} &
  \multicolumn{1}{c|}{\cellcolor[HTML]{C0C0C0}37} &
  2.1 &
  \multicolumn{1}{c|}{\cellcolor[HTML]{C0C0C0}34} &
  \multicolumn{1}{c|}{\cellcolor[HTML]{C0C0C0}36} &
  \multicolumn{1}{c|}{\cellcolor[HTML]{C0C0C0}38} &
  \multicolumn{1}{c|}{\cellcolor[HTML]{C0C0C0}39} &
  \multicolumn{1}{c|}{\cellcolor[HTML]{C0C0C0}39} &
  \multicolumn{1}{c|}{\cellcolor[HTML]{C0C0C0}38} &
  \multicolumn{1}{c|}{\cellcolor[HTML]{C0C0C0}36} &
  \multicolumn{1}{c|}{\cellcolor[HTML]{C0C0C0}-} &
  - \\ \cline{2-13} 
\rowcolor[HTML]{C0C0C0} 
\multirow{-2}{*}{\cellcolor[HTML]{C0C0C0}\begin{tabular}[c]{@{}c@{}}Channel \\ diameter ($\mu m$)\end{tabular}} &
  \multicolumn{1}{c|}{\cellcolor[HTML]{C0C0C0}VOI(4)} &
  \multicolumn{1}{c|}{\cellcolor[HTML]{C0C0C0}48} &
  3.4 &
  \multicolumn{1}{c|}{\cellcolor[HTML]{C0C0C0}42} &
  \multicolumn{1}{c|}{\cellcolor[HTML]{C0C0C0}44} &
  \multicolumn{1}{c|}{\cellcolor[HTML]{C0C0C0}47} &
  \multicolumn{1}{c|}{\cellcolor[HTML]{C0C0C0}49} &
  \multicolumn{1}{c|}{\cellcolor[HTML]{C0C0C0}49} &
  \multicolumn{1}{c|}{\cellcolor[HTML]{C0C0C0}50} &
  \multicolumn{1}{c|}{\cellcolor[HTML]{C0C0C0}50} &
  \multicolumn{1}{c|}{\cellcolor[HTML]{C0C0C0}51} &
  52 \\ \hline
\rowcolor[HTML]{EFEFEF} 
\cellcolor[HTML]{EFEFEF} &
  \multicolumn{1}{c|}{\cellcolor[HTML]{EFEFEF}VOI(3)} &
  \multicolumn{1}{c|}{\cellcolor[HTML]{EFEFEF}7.45} &
  0.8 &
  \multicolumn{1}{c|}{\cellcolor[HTML]{EFEFEF}7.51} &
  \multicolumn{1}{c|}{\cellcolor[HTML]{EFEFEF}7.55} &
  \multicolumn{1}{c|}{\cellcolor[HTML]{EFEFEF}7.50} &
  \multicolumn{1}{c|}{\cellcolor[HTML]{EFEFEF}8.03} &
  \multicolumn{1}{c|}{\cellcolor[HTML]{EFEFEF}8.18} &
  \multicolumn{1}{c|}{\cellcolor[HTML]{EFEFEF}7.04} &
  \multicolumn{1}{c|}{\cellcolor[HTML]{EFEFEF}6.35} &
  \multicolumn{1}{c|}{\cellcolor[HTML]{EFEFEF}-} &
  - \\ \cline{2-13} 
\rowcolor[HTML]{EFEFEF} 
\multirow{-2}{*}{\cellcolor[HTML]{EFEFEF}Connectivity} &
  \multicolumn{1}{c|}{\cellcolor[HTML]{EFEFEF}VOI(4)} &
  \multicolumn{1}{c|}{\cellcolor[HTML]{EFEFEF}10.52} &
  1.0 &
  \multicolumn{1}{c|}{\cellcolor[HTML]{EFEFEF}9.42} &
  \multicolumn{1}{c|}{\cellcolor[HTML]{EFEFEF}10.06} &
  \multicolumn{1}{c|}{\cellcolor[HTML]{EFEFEF}10.56} &
  \multicolumn{1}{c|}{\cellcolor[HTML]{EFEFEF}11.02} &
  \multicolumn{1}{c|}{\cellcolor[HTML]{EFEFEF}10.80} &
  \multicolumn{1}{c|}{\cellcolor[HTML]{EFEFEF}10.87} &
  \multicolumn{1}{c|}{\cellcolor[HTML]{EFEFEF}11.03} &
  \multicolumn{1}{c|}{\cellcolor[HTML]{EFEFEF}10.66} &
  10.27 \\ \hline
\rowcolor[HTML]{C0C0C0} 
\cellcolor[HTML]{C0C0C0} &
  \multicolumn{1}{c|}{\cellcolor[HTML]{C0C0C0}VOI(3)} &
  \multicolumn{1}{c|}{\cellcolor[HTML]{C0C0C0}-} &
  - &
  \multicolumn{1}{c|}{\cellcolor[HTML]{C0C0C0}1.30} &
  \multicolumn{1}{c|}{\cellcolor[HTML]{C0C0C0}1.32} &
  \multicolumn{1}{c|}{\cellcolor[HTML]{C0C0C0}1.41} &
  \multicolumn{1}{c|}{\cellcolor[HTML]{C0C0C0}1.18} &
  \multicolumn{1}{c|}{\cellcolor[HTML]{C0C0C0}1.38} &
  \multicolumn{1}{c|}{\cellcolor[HTML]{C0C0C0}1.15} &
  \multicolumn{1}{c|}{\cellcolor[HTML]{C0C0C0}1.22} &
  \multicolumn{1}{c|}{\cellcolor[HTML]{C0C0C0}-} &
  - \\ \cline{2-13} 
\rowcolor[HTML]{C0C0C0} 
\multirow{-2}{*}{\cellcolor[HTML]{C0C0C0}Tortuosity} &
  \multicolumn{1}{c|}{\cellcolor[HTML]{C0C0C0}VOI(4)} &
  \multicolumn{1}{c|}{\cellcolor[HTML]{C0C0C0}-} &
  - &
  \multicolumn{1}{c|}{\cellcolor[HTML]{C0C0C0}1.18} &
  \multicolumn{1}{c|}{\cellcolor[HTML]{C0C0C0}1.19} &
  \multicolumn{1}{c|}{\cellcolor[HTML]{C0C0C0}1.20} &
  \multicolumn{1}{c|}{\cellcolor[HTML]{C0C0C0}1.16} &
  \multicolumn{1}{c|}{\cellcolor[HTML]{C0C0C0}1.16} &
  \multicolumn{1}{c|}{\cellcolor[HTML]{C0C0C0}1.18} &
  \multicolumn{1}{c|}{\cellcolor[HTML]{C0C0C0}1.18} &
  \multicolumn{1}{c|}{\cellcolor[HTML]{C0C0C0}1.17} &
  1.20 \\ \hline
\end{tabular}%
}
\caption{Average architectural parameters in the sections shown in Fig.\ref{fig:PorosityvsL}D}
\label{table:archi}
\end{table}

\subsection{Velocity and pressure inside the meniscal channels}

\begin{figure}
    \centering
    \begin{subfigure}[b]{0.49\textwidth}
        \centering
        \includegraphics[width=\textwidth]{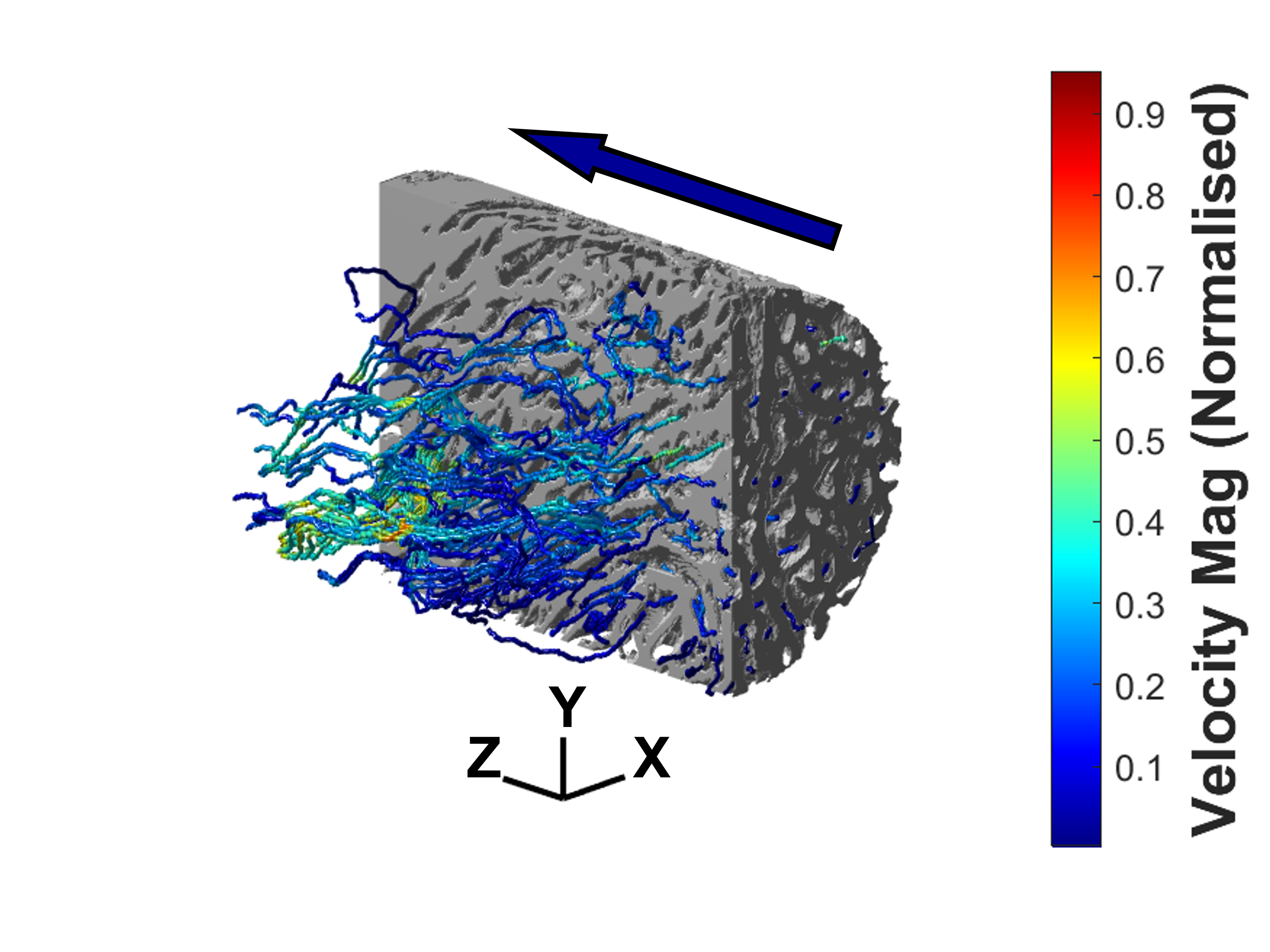}
        \caption{VOI3 Velocity Streamlines at 0.1mm/s}
        \label{fig:my_label}
    \end{subfigure}
    \hfill\begin{subfigure}[b]{0.49\textwidth}
        \centering
        \includegraphics[width=\textwidth]{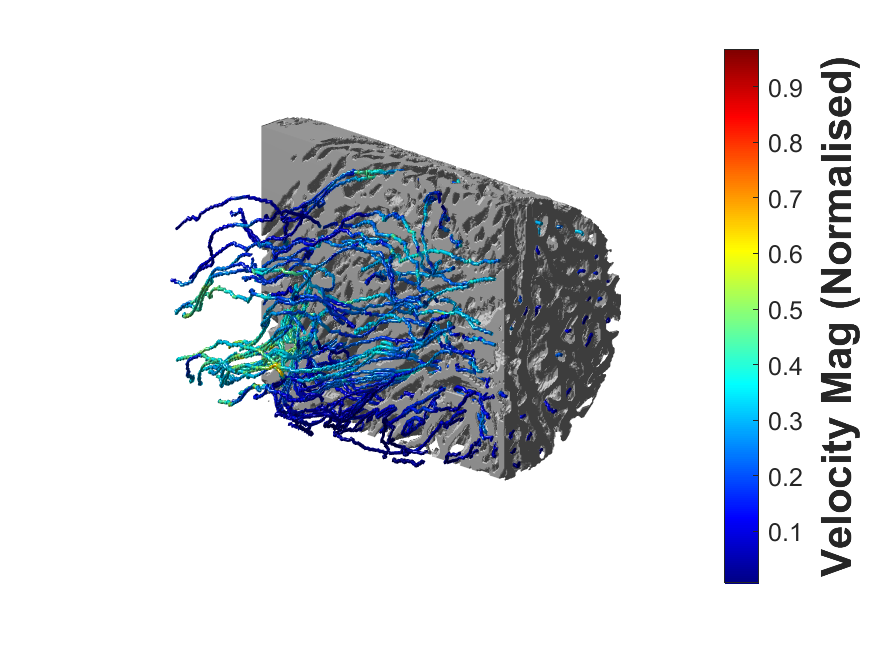}
         \caption{VOI3 Velocity Streamlines at 1.64025m/s}
        \label{fig:my_label}
    \end{subfigure}
     \hfill\begin{subfigure}[b]{0.49\textwidth}
        \centering
        \includegraphics[width=\textwidth]{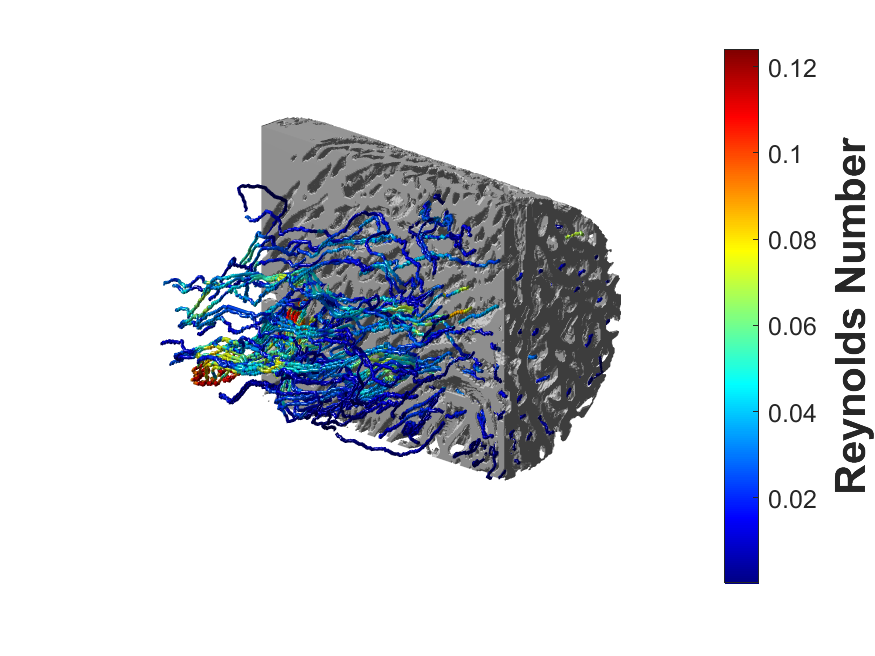}
         \caption{VOI3 Re Number Variation at 0.1mm/s}
        \label{fig:my_label}
    \end{subfigure}
      \hfill\begin{subfigure}[b]{0.49\textwidth}
        \centering
        \includegraphics[width=\textwidth]{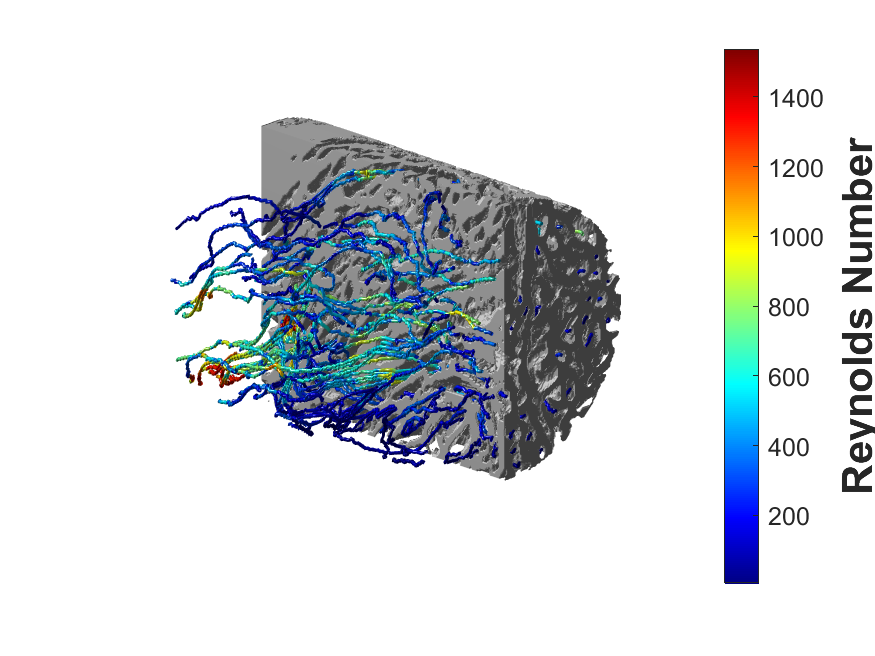}
         \caption{VOI3 Re Number Variation at 1.64025m/s}
        \label{fig:my_label}
    \end{subfigure}
 \hfill\begin{subfigure}[b]{0.49\textwidth}
        \centering
        \includegraphics[width=\textwidth]{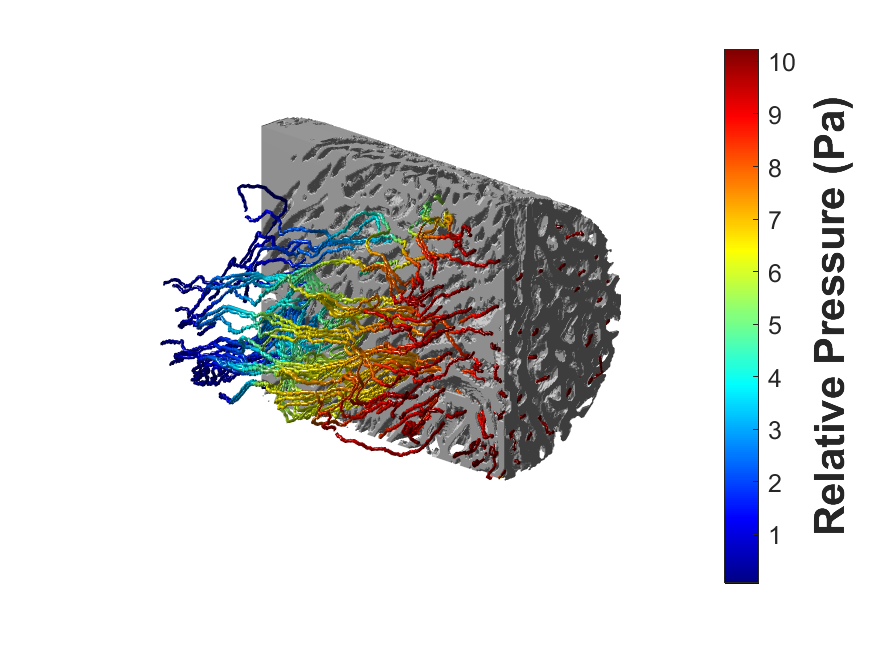}
         \caption{VOI3 Relative pressure at 0.1mm/s}
        \label{fig:my_label}
    \end{subfigure}
     \hfill\begin{subfigure}[b]{0.49\textwidth}
        \centering
        \includegraphics[width=\textwidth]{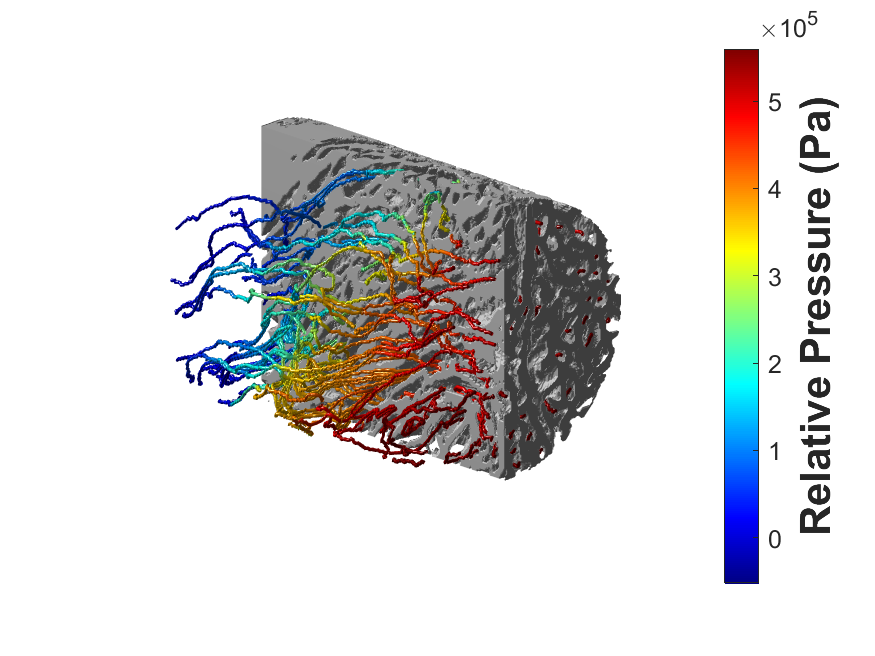}
         \caption{VOI3 Relative pressure at 1.64025m/s}
        \label{fig:my_label}
    \end{subfigure}
    \caption{ VOI(3) streamlines results. (a) Normalised Velocity Magnitude Streamlines normalised by the reported maximum velocity (Vmax = 23.84 and 2.1$\times 10^{-4} m/s$ @ 1.6$m/s$ and @ 0.1$mm/s$ respectively, released from the center of each pore at z=0 of the sample (example shown in Fig.1 D) for the (lowest) inlet velocity of 0.1mm/s. (b) Streamlines velocity magnitude for the highest inlet velocity of 1.6m/s. (c) Re number along streamlines for the (lowest) inlet velocity of 0.1mm/s. (d) Re number along streamlines for the (highest) inlet velocity of 1.6m/s. (e) Streamlines relative pressure, i.e. pressure inside channels for the case of inlet velocity of 0.1mm/s. (f) Streamlines relative pressure, i.e. pressure inside channels for the case of inlet velocity of 1.6m/s.}  
    \label{Figure: VOI Data}
\end{figure}

\begin{figure}
    \centering
    \begin{subfigure}[b]{0.49\textwidth}
        \centering
        \includegraphics[width=\textwidth]{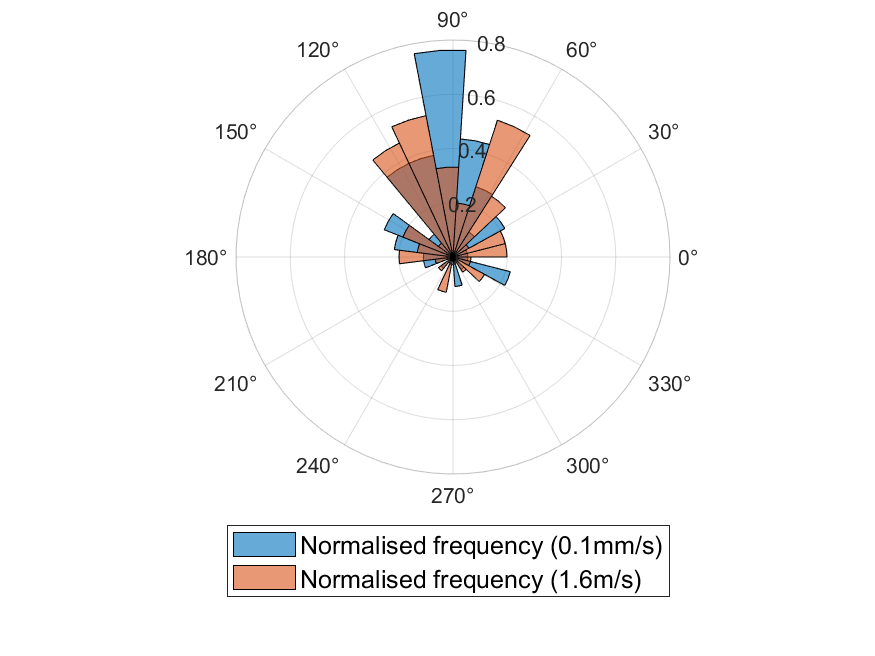}
        \caption{VOI3 Flow Angularity Distribution}
        \label{fig:my_label}
    \end{subfigure}
    \hfill\begin{subfigure}[b]{0.49\textwidth}
        \centering
        \includegraphics[width=\textwidth]{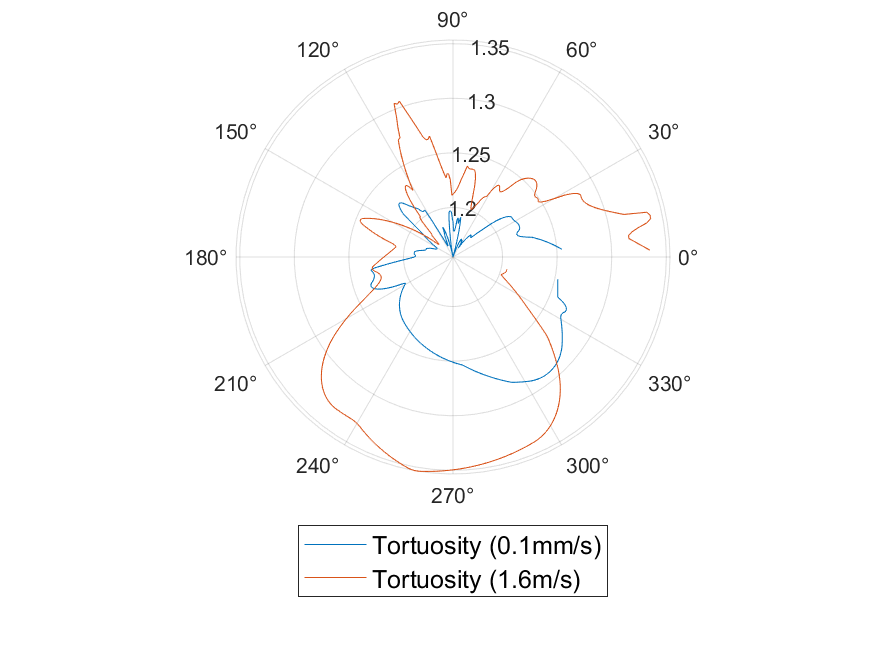}
         \caption{VOI3 Tortuosity Angularity Polar Plot}
        \label{fig:my_label}
    \end{subfigure}
     \hfill\begin{subfigure}[b]{0.49\textwidth}
        \centering
        \includegraphics[width=\textwidth]{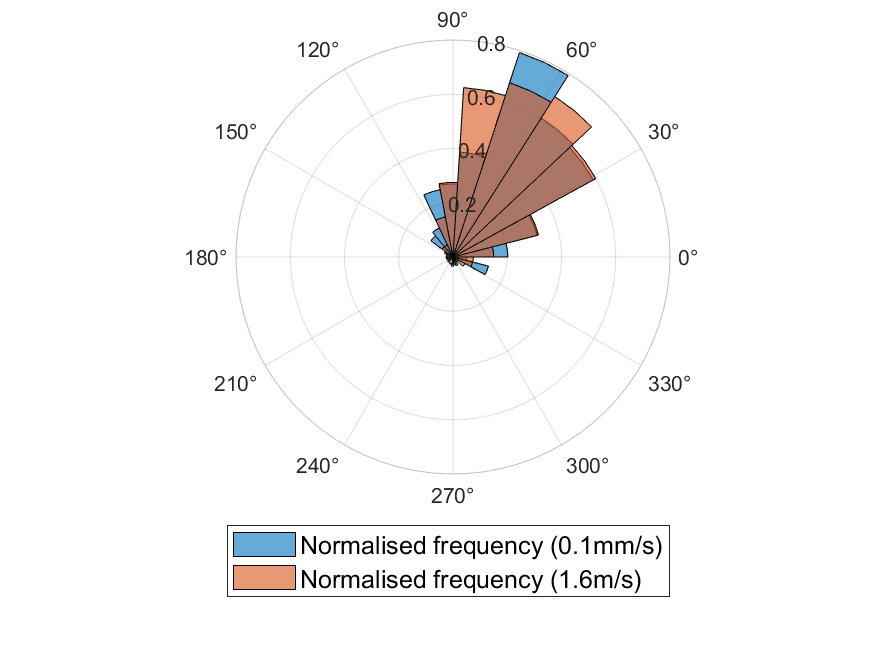}
         \caption{VOI4 Flow Angularity Distribution}
        \label{fig:my_label}
    \end{subfigure}
      \hfill\begin{subfigure}[b]{0.49\textwidth}
        \centering
        \includegraphics[width=\textwidth]{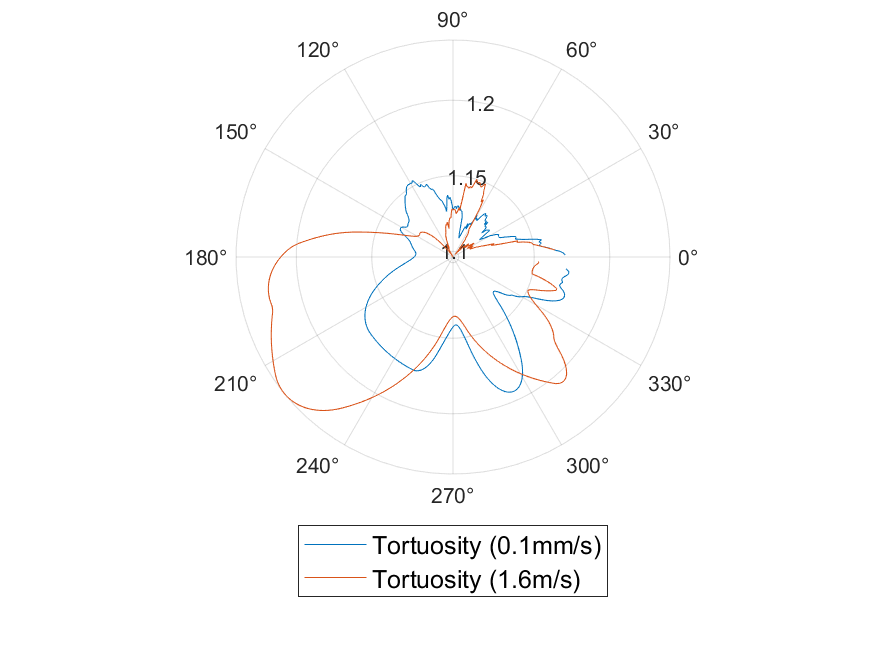}
         \caption{VOI4 Tortuosity Angularity Polar Plot}
        \label{fig:my_label}
    \end{subfigure}
    \caption{Orientation analysis of tortuosity. (a) Polar histogram of VOI(3) streamlines, distribution parameters at 0.1: $\mu = 99^{\circ}, \kappa = 1.34$ and at 1.6: $\mu = 92^{\circ}, \kappa = 0.88$ (b) Polar plot of VOI(3) streamline orientation and associated tortuosity (c) Polar histogram of VOI(4) streamlines, distribution parameters at 0.1: $\mu = 57^{\circ}, \kappa = 2.02$ and at 1.6: $\mu = 63^{\circ}, \kappa = 1.72$ (d) Polar plot of VOI(4) streamline orientation and associated tortuosity.}
    \label{Figure: VOI Tort and Anglularity Data}
\end{figure}

Figure \ref{Figure: VOI Data} provides a visualisation of fluid streamlines within the flow domain of VOI(3), along with the associated parameters at these locations. The normalised velocity magnitude can be seen in Fig. \ref{Figure: VOI Data} (a, b), Reynolds number in Fig. \ref{Figure: VOI Data} (c, d) and relative pressure in Fig. \ref{Figure: VOI Data} (e, f), for both the lowest and highest inlet velocities, 0.1$mm/s$ and $1.64m/s$. One immediate distinction that is visually derived from the two sets of streamlines is that the streams differ between the two inlet velocities, despite being released from identical locations. Quantifying and understanding how these paths evolve at different inlet velocities is crucial, as these could be seen as analogous to the behaviour of the tissue under various strain-rate loading conditions. Low inlet velocities may possibly be representative of low strain-rate movements, such as static loading due to standing, while high velocities may be more conducive to jumping from a height or running. The data presented in Figure \ref{Figure: VOI Tort and Anglularity Data} aims to quantify the discrepancies in these streamlines. Firstly, Fig.\ref{Figure: VOI Tort and Anglularity Data} (a,c) present polar histograms representing the distribution of stream trajectory in the XY plane, demonstrating a clear preferential direction of the fluid paths. These distributions have been modelled as Von-Mises distributions to determine the preferential direction and the 'kappa' concentration parameter
\cite{MARQUEZ}, which provides an indication of how much the flow path deviates from the primary orientation. Kappa is a non-dimensional value that varies from 0-$\inf$, with 0 indicating no preferential orientation and $\inf$ indicating perfect alignment with no variation \cite{Schriefl2012}. The preferential directions are ~93.5$^{\circ}$ \& 56$^{\circ}$ for VOI(3) and VOI(4) respectively. While these preferential directions only vary by 1-5$^{\circ}$ between the low and high inlet velocities in both samples, the Kappa value decreases by an average of 0.4 at higher velocities. This illustrates that at higher velocities the preferential direction is (broadly speaking) unchanged, the fluid is dispersed less uniformly throughout the flow domain. Along with scattering, Fig.\ref{Figure: VOI Tort and Anglularity Data}(b,d) also demonstrates that fluid paths are generally more tortuous at higher velocities, with the average tortuosity of a fluid path being 10.4\% greater in the 1.6$m/s$ simulations, although it is worth noting that this value did vary considerably between the two samples.

\par

Figure \ref{Figure: VOI Data} shows that the largest velocity magnitude occurs towards the end of the VOI(3) sample, along the flow direction. This can be attributed to the drop in porosity seen at the end of the sample, where the tissue structure seemingly transitions rapidly from one form to another. This rapid decrease in porosity in essence forces the remaining fluid through a declining set of fluid passageways, leading to the localised high-velocity, and hence Re, seen at the end of the sample. Re numbers range from 0.01-0.1 for 0.1$mm/s$, meanwhile at 1.64$m/s$ Re is seen to reach peaks of 1400, showing that some channels are in transition to a turbulent regime towards the end of the sample.  These are located where the flow area decreases in VOI(3) by 17\% over a length of 0.8mm, this is a location at which the flow angle is inclined relative to the axial, z-axis, forcing flow towards the wall boundary of the sample and generating flow convergence. Figure \ref{Figure: VOI Data} c-d highlights the alignment of locations in which elevated Re occurs for both inlet velocities. Fig. \ref{Figure: VOI Data} (e,f) present the relative pressure within the sample for both inlet velocities, this pressure is relative to the absolute reference pressure of 101325 Pa. The preferential direction of the channels, ~93.5$^{\circ}$ for VOI(3), impacts the pressure drop between the start of the sample and the end along the flow direction (Z - axis), varying from 6.49$\times10^5 Pa$ to 9.85$\times10^4 Pa$. The pressure is observed to vary radially (1.6mm along the VOI) from 5.96$\times10^5 Pa$ to 2.08$\times10^5 Pa$ at 1.64$m/s$ inlet velocity. Simulations of VOI(4) showing similar results are presenting in the supporting material Fig. S3, despite the absence of sudden changes in microstructure, and hence porosity, observed in VOI(3). The fluid flow is oriented at ~56$^{\circ}$ for the VOI(4) sample and the drop in pressure varies from 5.58$\times10^5 Pa$ to 9.68$\times10^5 Pa$ in the axial z-direction and from 4.22$\times10^5 Pa$ to 1.43$\times10^5 Pa$ radially (y-direction as shown in Fig. \ref{Figure: VOI Data} a ) across the VOI (1.8mm along VOI(4) the location where the pressure gradient was observed to be largest from simulations) at an inlet velocity of 1.64$m/s$.  Data indicates the pressure range along the sample length is comparable for both VOI(3) and VOI(4) (see Fig. S3 in supporting information), despite the differences in morphology. It is observed that in both samples, the preferential flow direction produces flow impingement upon the wall boundary of the sample, this is thought to contribute to the variation in the radial pressure for each sample, a phenomenon that would occur in experimental Darcy configurations of the samples and which implies the sample extraction orientation can be influential on data obtained from such work. 

\par

\begin{figure}
    \centering
    \begin{subfigure}[b]{0.49\textwidth}
      \centering
         \includegraphics[width=\textwidth]{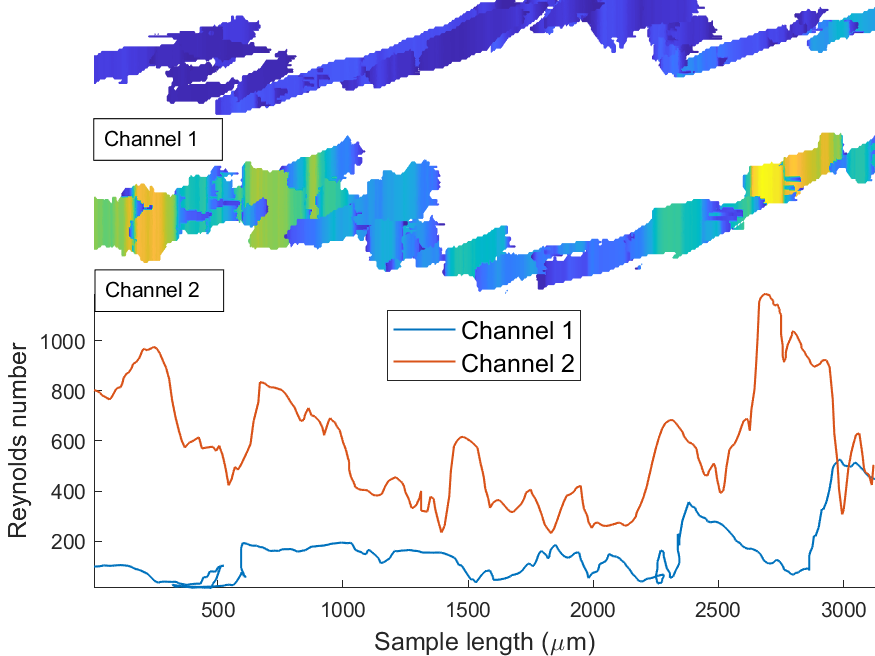}
         \caption{Morphology and Re of the two meniscal channels in VOI(3)}
        \label{fig:my_label}
    \end{subfigure}
     \hfill
    \begin{subfigure}[b]{0.49\textwidth}
      \centering
         \includegraphics[width=\textwidth]{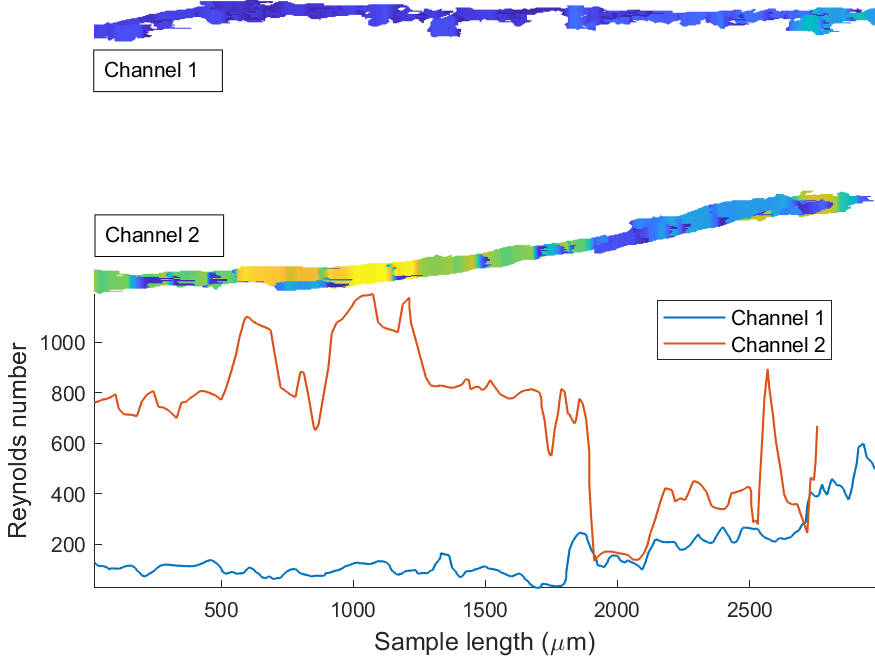}
         \caption{Morphology and Re of the two meniscal channels in VOI(4)}
        \label{fig:my_label}
    \end{subfigure}
     \hfill
     \begin{subfigure}[b]{0.49\textwidth}
        \centering
        \includegraphics[width=\textwidth]{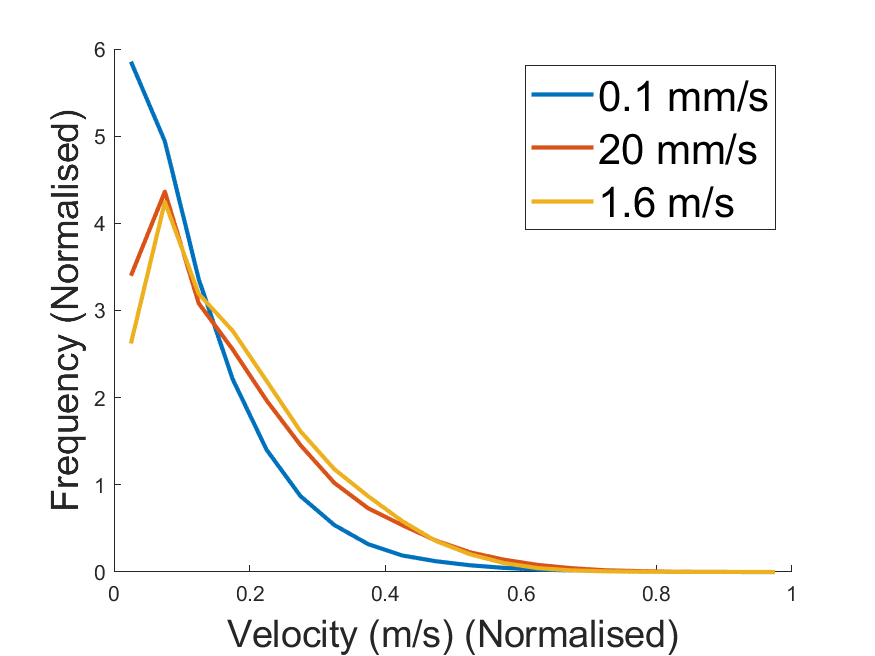}
        \caption{Velocity distribution of VOI(3)}
        \label{fig:my_label}
    \end{subfigure}
    \hfill
    \begin{subfigure}[b]{0.49\textwidth}
        \centering
        \includegraphics[width=\textwidth]{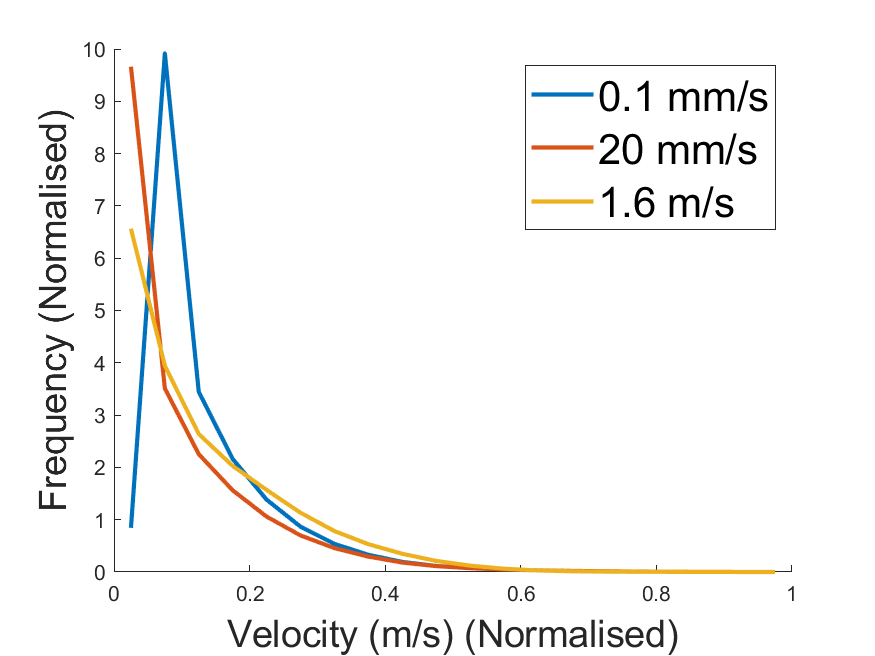}
        \caption{Velocity distribution of VOI(4)}
        \label{fig:my_label}
    \end{subfigure}
     \hfill
     \begin{subfigure}[b]{0.49\textwidth}
        \centering
        \includegraphics[width=\textwidth]{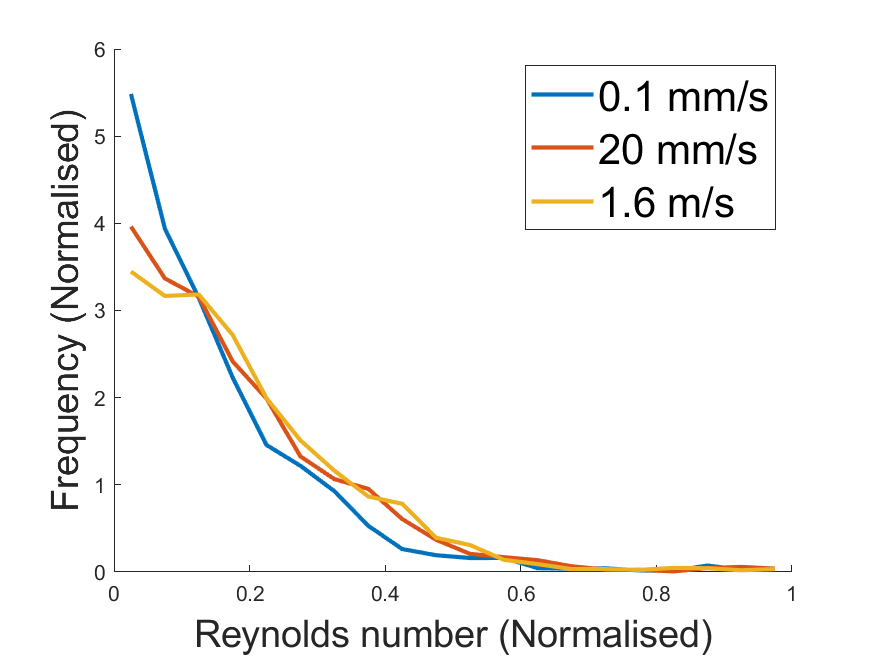}
         \caption{Re distribution of VOI(3)}
        \label{fig:my_label}
    \end{subfigure}
    \hfill
    \begin{subfigure}[b]{0.49\textwidth}
        \centering
        \includegraphics[width=\textwidth]{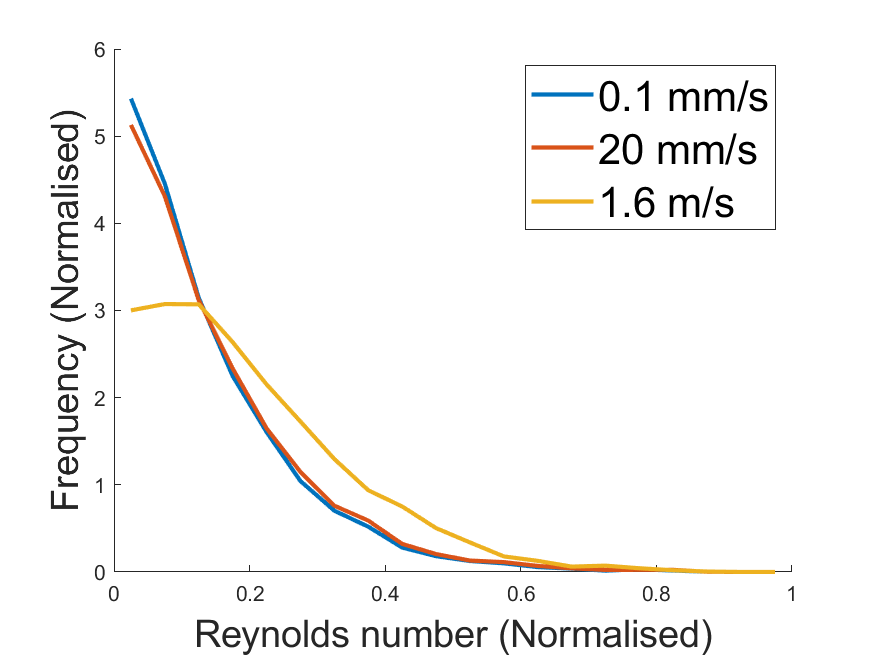}
        \caption{Re distribution of VOI(4)}
        \label{fig:my_label}
    \end{subfigure}
    \caption{(a)Morphology of the two meniscal channels with the lowest and highest average Re in VOI(3), for an inlet velocity of 1.6$m/s$. The variation in Re Channel 1 (lowest) and Channel 2 (highest) is plotted below. (b) Morphology of the two meniscal channels with the lowest and highest average Re in VOI(4), for an inlet velocity of 1.6$m/s$. The variation in Re Channel 1 (lowest) and Channel 2 (highest) is plotted below. (c) Normalised  velocity distribution in VOI(3) with respect to $V_{max}$ for 0.1$mm/s$: 0.0015, 20$mm/s$: 0.32 \& 1.6$m/s$: 21.81$m/s$. (d) Normalised  velocity distribution in VOI(3) with respect to $V_{max}$ for 0.1$mm/s$: 0.0015, 20$mm/s$: 0.27 \& 1.6$m/s$: 16.57$m/s$. (e) Normalised  Re distribution in VOI(3) with respect to $Re_{max}$ for 0.1$mm/s$: 0.12, 20$mm/s$: 23.88 \& 1.6$m/s$: 1357.64. (f) Normalised  Re distribution in VOI(4) with respect to $Re_{max}$ for 0.1$mm/s$: 0.12, 20$mm/s$: 24.34 \& 1.6$m/s$: 1471.37.}
    \label{Figure: VOI Velocity and Re Distribution}
\end{figure}

It has been noted that Re varies considerably along streamlines, up to 400\%. The combined CFD-IA methodology presented in this work, of mapping individual streamlines within channels grants the opportunity 'follow' the evolution of Re within channels. To demonstrate two varied examples of this, the evolution of Re within two channels of the VOI(3) and VOI(4) samples have been shown in Fig.\ref{Figure: VOI Velocity and Re Distribution}(a-b). Both figures present two cases of streamlines with unique paths and channels, 'Channel 1' in each case represents the streamline with the lowest average Re, while 'Channel 2' presents the highest. Channel 1 in both samples varies between 10 to 575 on average, with peak values reached around sample length 2750 and 3000 $\mu$m. The Re in channel 2 varies between 78 to 1200 on average, with peak values of Re observed at various locations in either sample.

\par

\begin{figure}[htbp]
    \centering
    \begin{subfigure}[b]{0.3\textwidth}
        \centering
        \includegraphics[width=\textwidth]{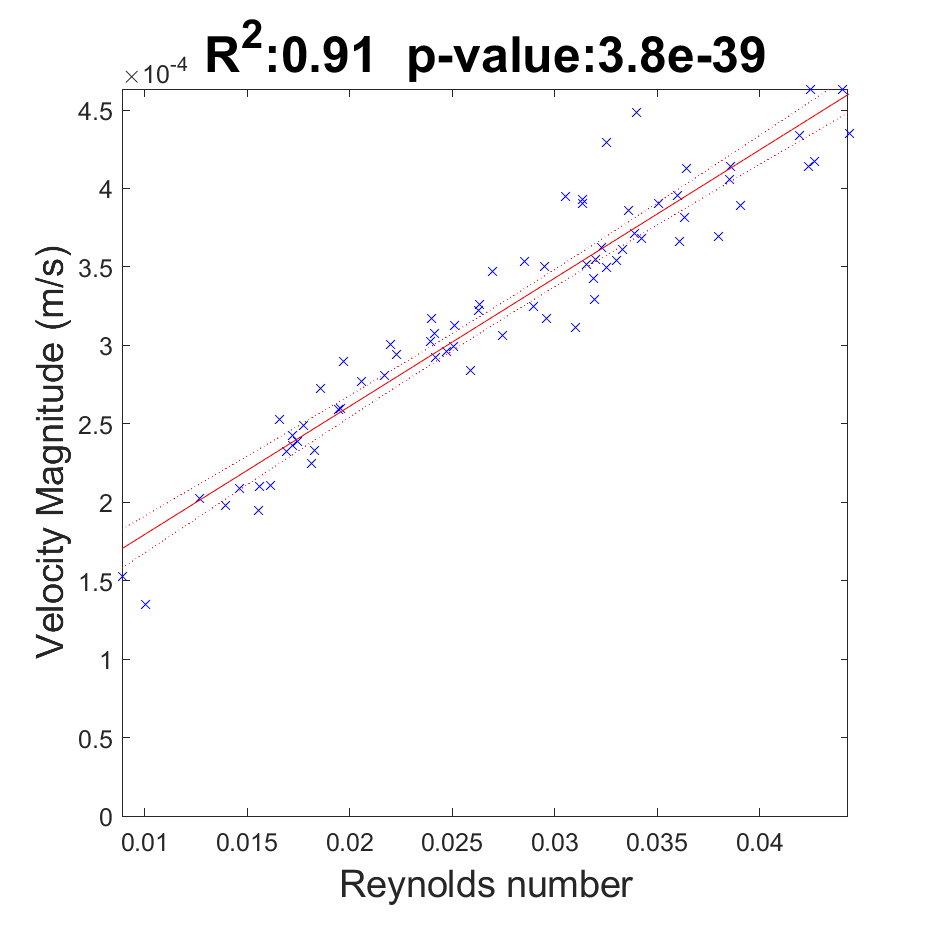}
        \caption{VOI3 Velocity-Re at 0.1mm/s}
        \label{fig:my_label}
    \end{subfigure}
    \hfill\begin{subfigure}[b]{0.3\textwidth}
        \centering
        \includegraphics[width=\textwidth]{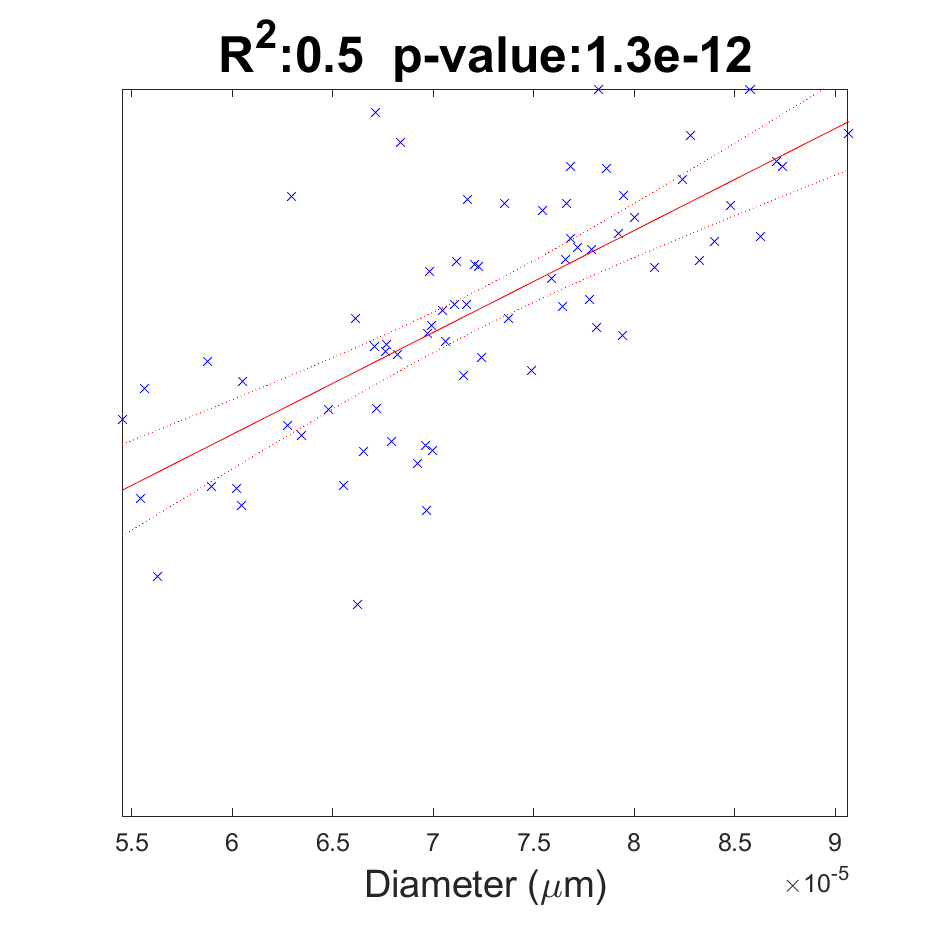}
         \caption{VOI3 Velocity-Diameter at 0.1mm/s}
        \label{fig:my_label}
    \end{subfigure}
     \hfill\begin{subfigure}[b]{0.3\textwidth}
        \centering
        \includegraphics[width=\textwidth]{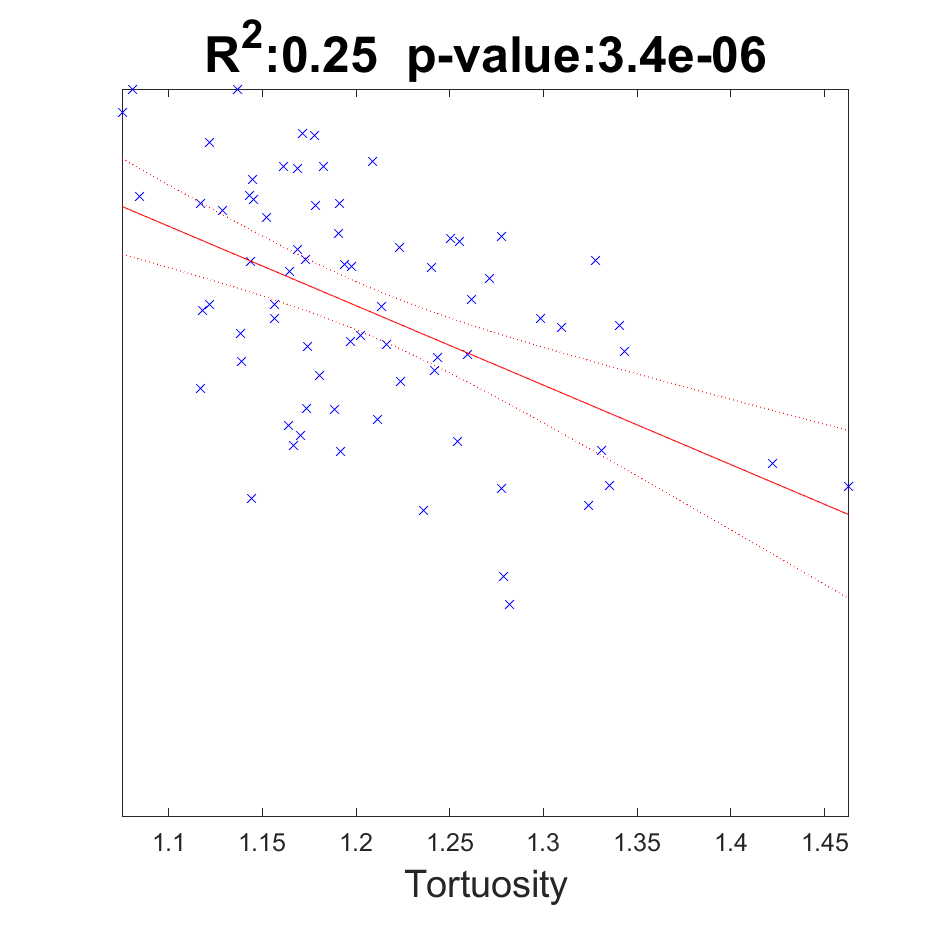}
         \caption{VOI3 Velocity-Tortuosity 0.1mm/s}
        \label{fig:my_label}
    \end{subfigure}
      \hfill\begin{subfigure}[b]{0.3\textwidth}
        \centering
        \includegraphics[width=\textwidth]{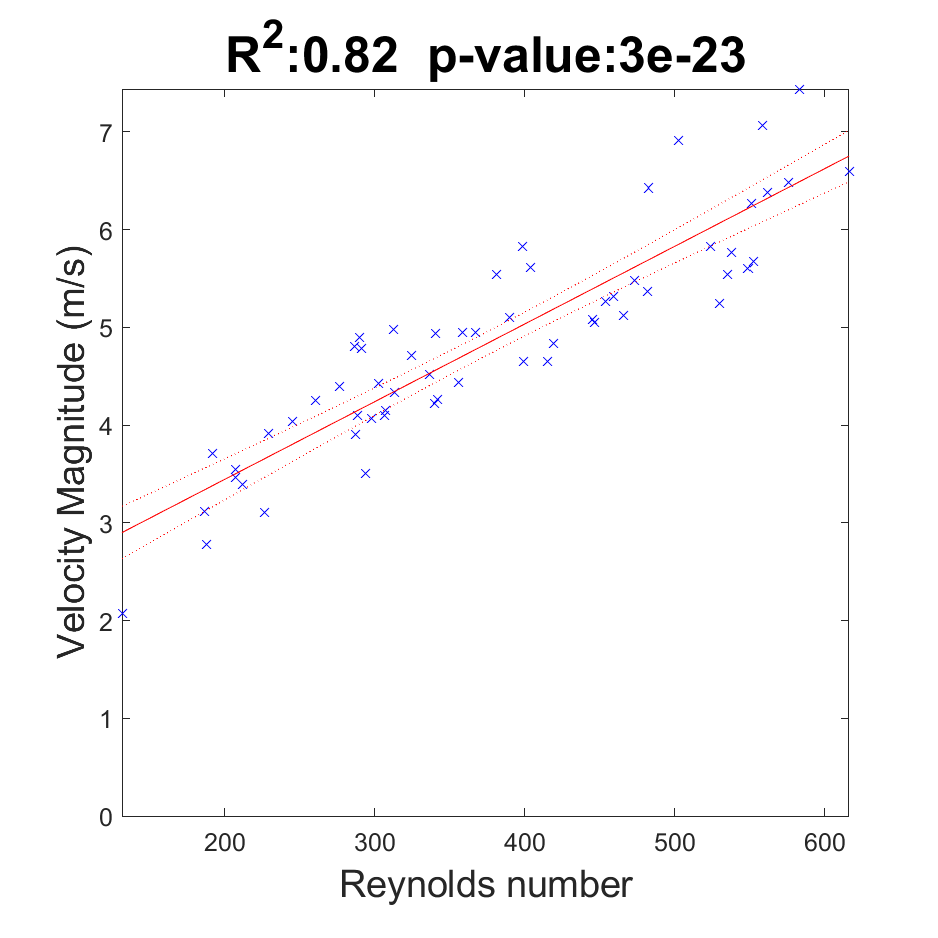}
         \caption{VOI3 Velocity-Re at 1.6m/s}
        \label{fig:my_label}
    \end{subfigure}
       \hfill\begin{subfigure}[b]{0.3\textwidth}
        \centering
        \includegraphics[width=\textwidth]{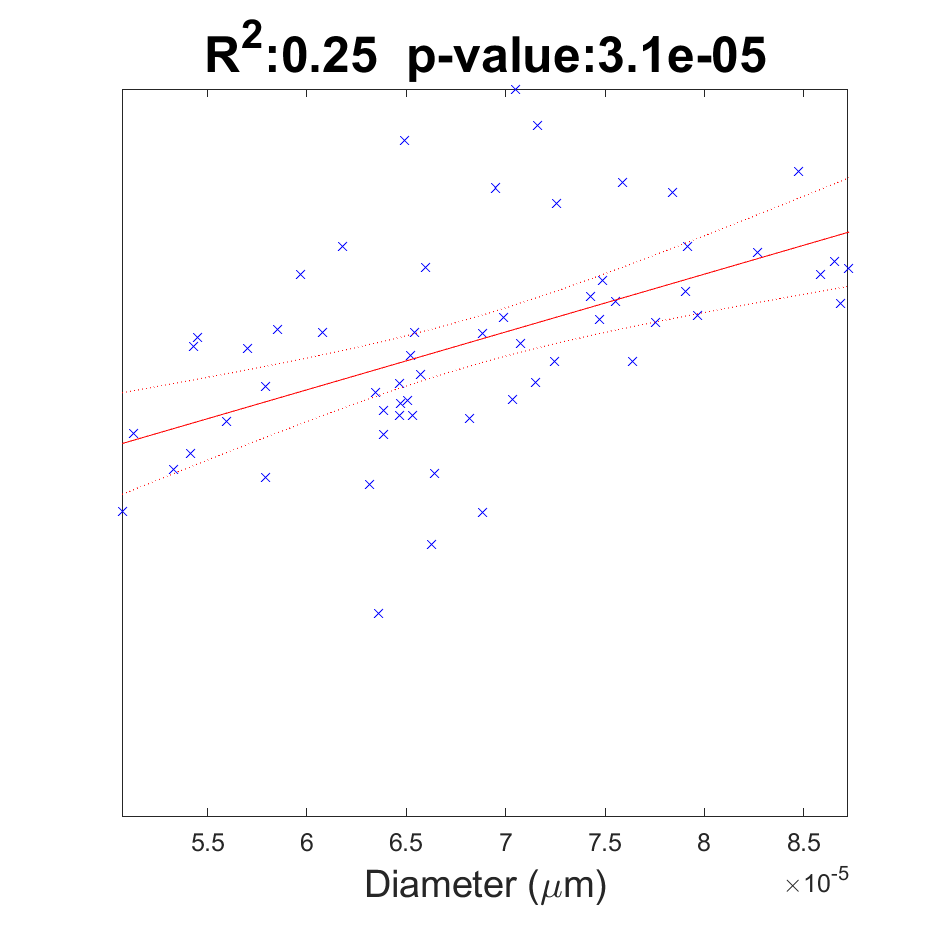}
         \caption{VOI3 Velocity-Diameter at 1.6m/s}
        \label{fig:my_label}
    \end{subfigure}
     \hfill\begin{subfigure}[b]{0.3\textwidth}
        \centering
        \includegraphics[width=\textwidth]{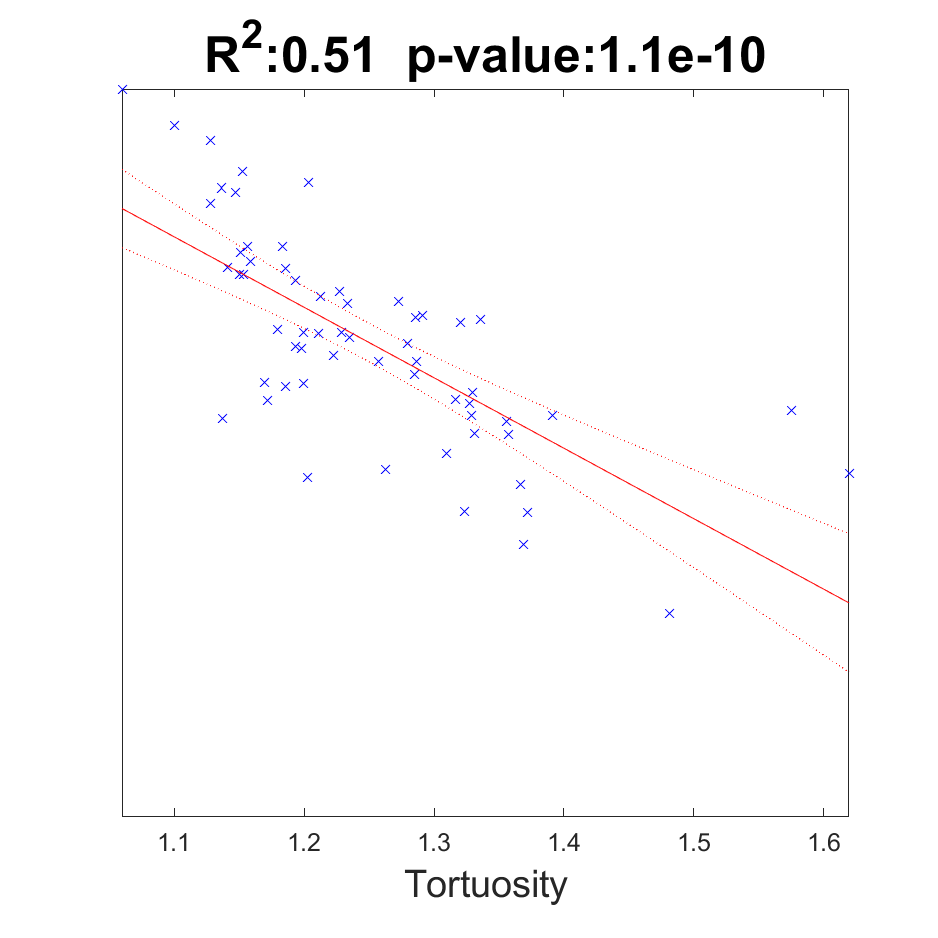}
         \caption{VOI3 Velocity-Tortuosity at 1.6m/s}
        \label{fig:my_label}
    \end{subfigure}
        \hfill\begin{subfigure}[b]{0.3\textwidth}
        \centering
        \includegraphics[width=\textwidth]{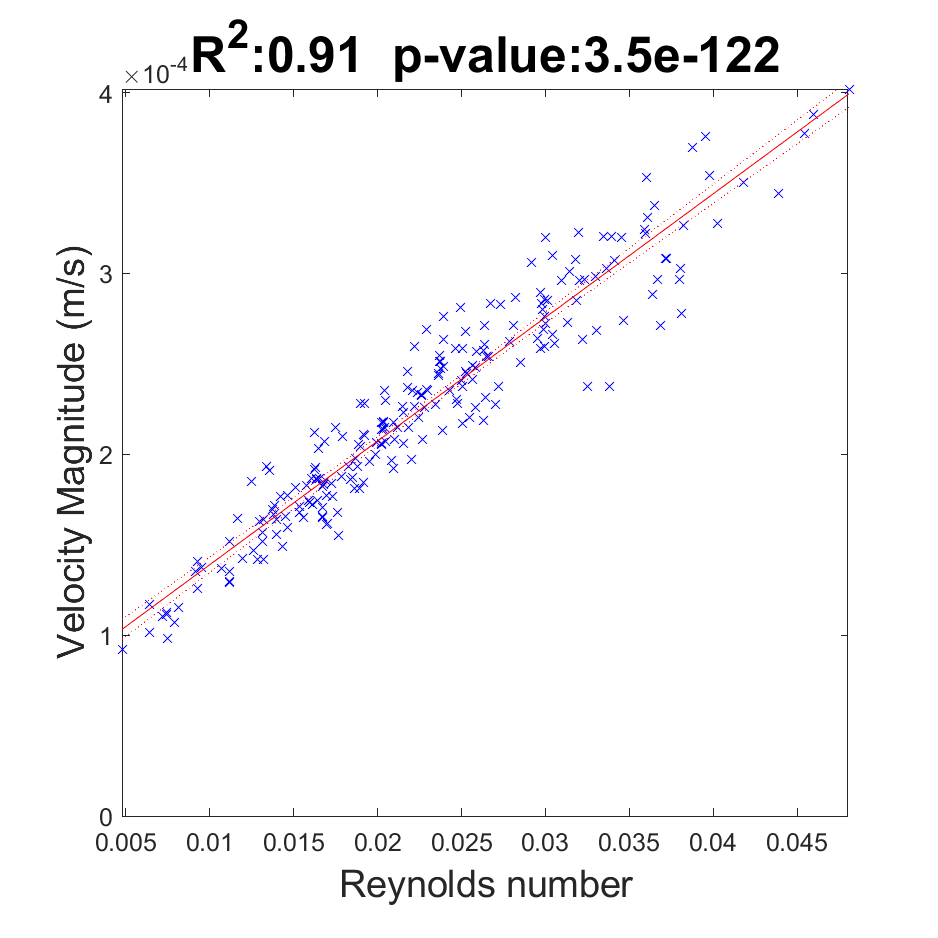}
         \caption{VOI4 Velocity-Diameter at 0.1mm/s}
        \label{fig:my_label}
    \end{subfigure}
     \hfill\begin{subfigure}[b]{0.3\textwidth}
        \centering
        \includegraphics[width=\textwidth]{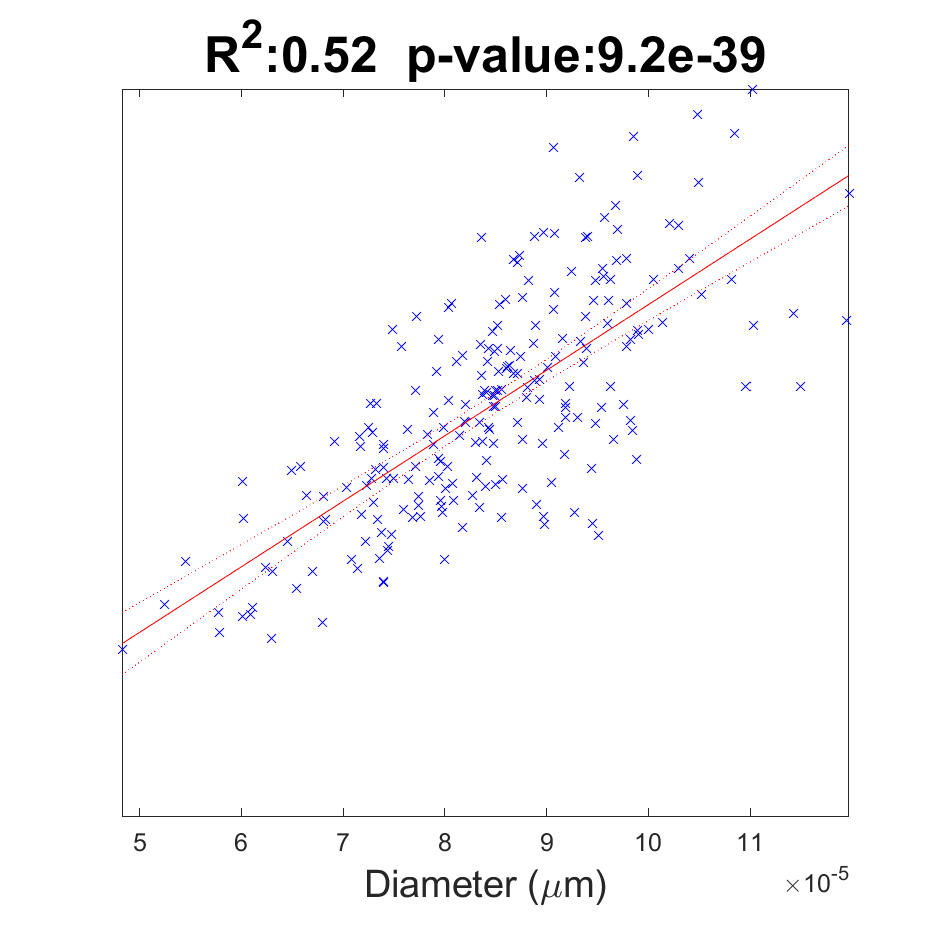}
         \caption{VOI4 Velocity-Diameter at 0.1mm/s}
        \label{fig:my_label}
    \end{subfigure}
      \hfill\begin{subfigure}[b]{0.3\textwidth}
        \centering
        \includegraphics[width=\textwidth]{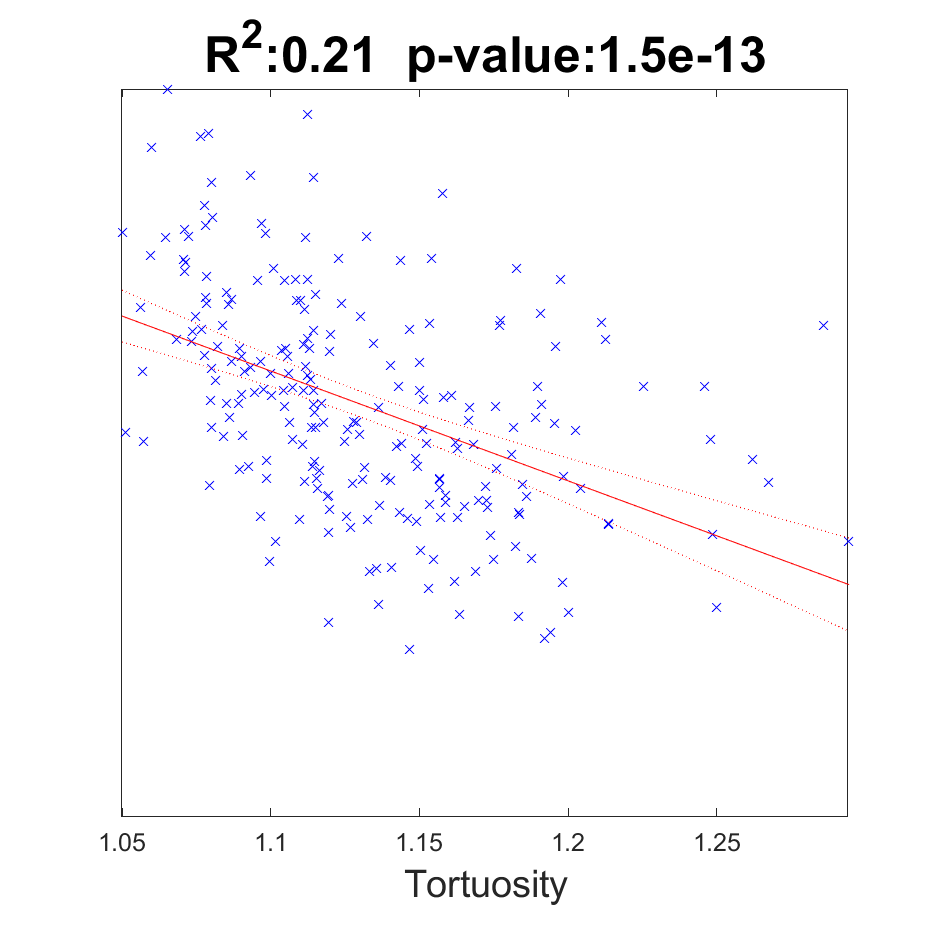}
         \caption{VOI4 Velocity-Tortuosity at 0.1mm/s}
        \label{fig:my_label}
    \end{subfigure}
    \hfill\begin{subfigure}[b]{0.3\textwidth}
        \centering
        \includegraphics[width=\textwidth]{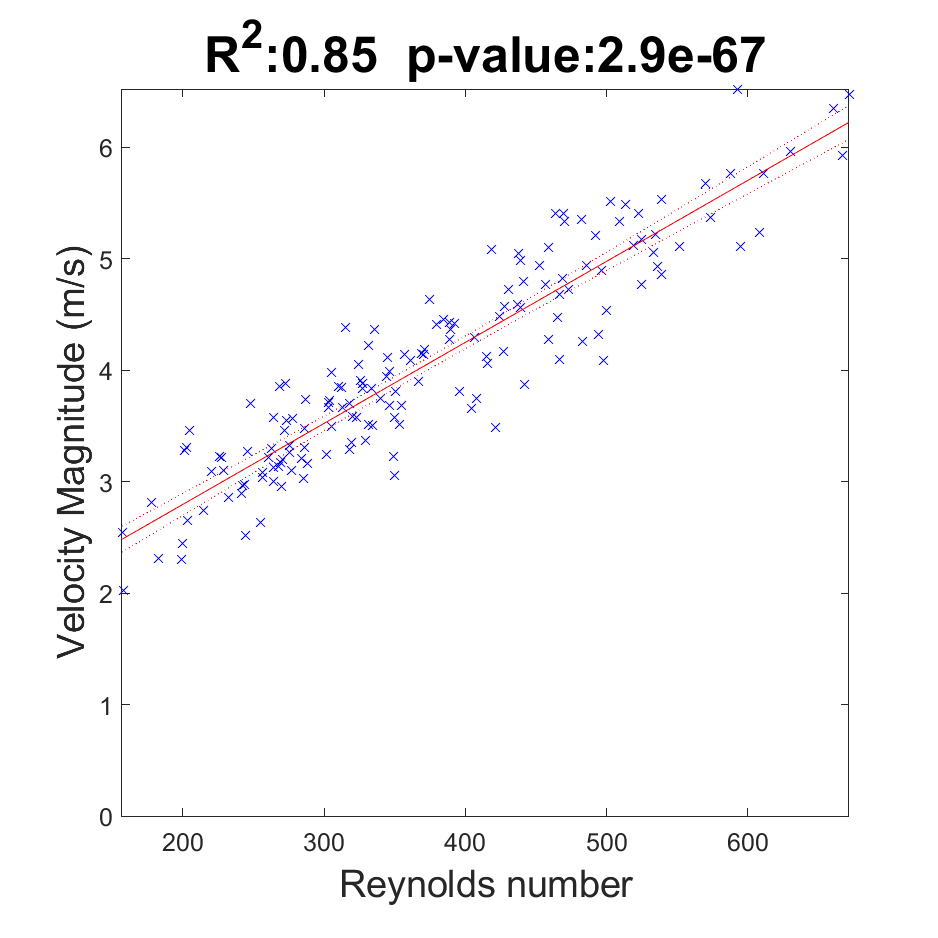}
         \caption{VOI4 Velocity-Re at 1.6m/s}
        \label{fig:my_label}
    \end{subfigure}
       \hfill\begin{subfigure}[b]{0.3\textwidth}
        \centering
        \includegraphics[width=\textwidth]{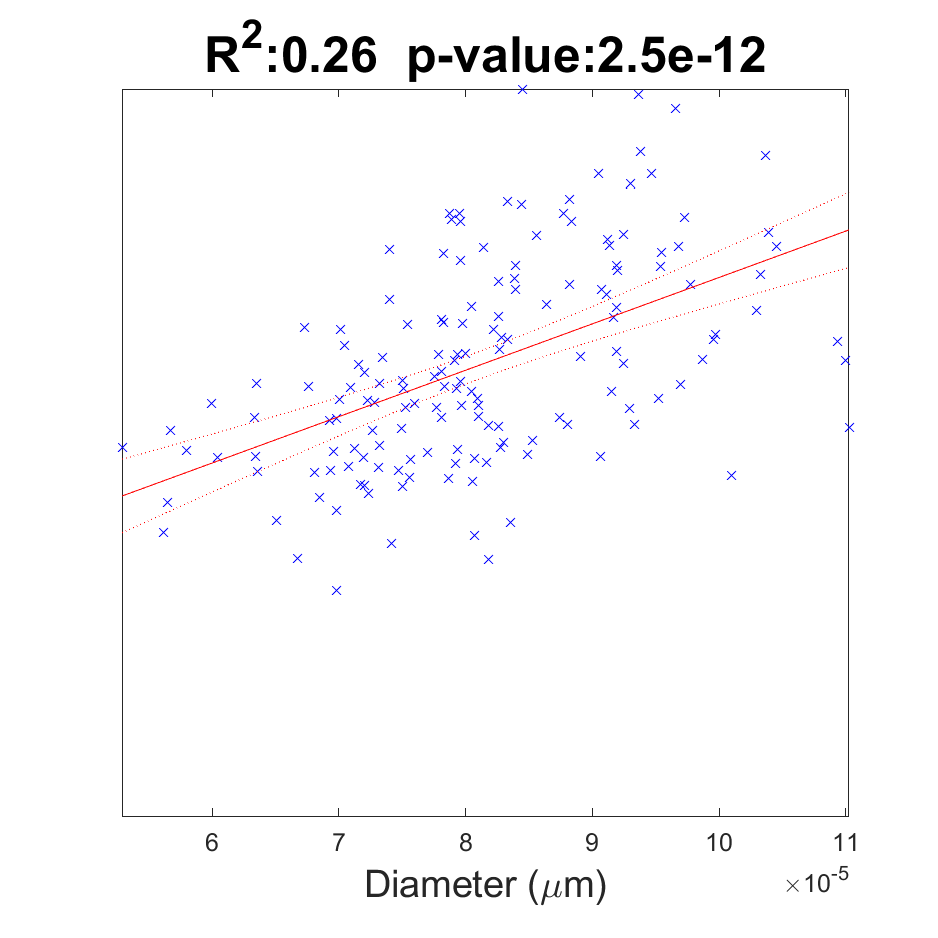}
         \caption{VOI4 Velocity-Diameter at 1.6m/s}
        \label{fig:my_label}
    \end{subfigure}
     \hfill\begin{subfigure}[b]{0.3\textwidth}
        \centering
        \includegraphics[width=\textwidth]{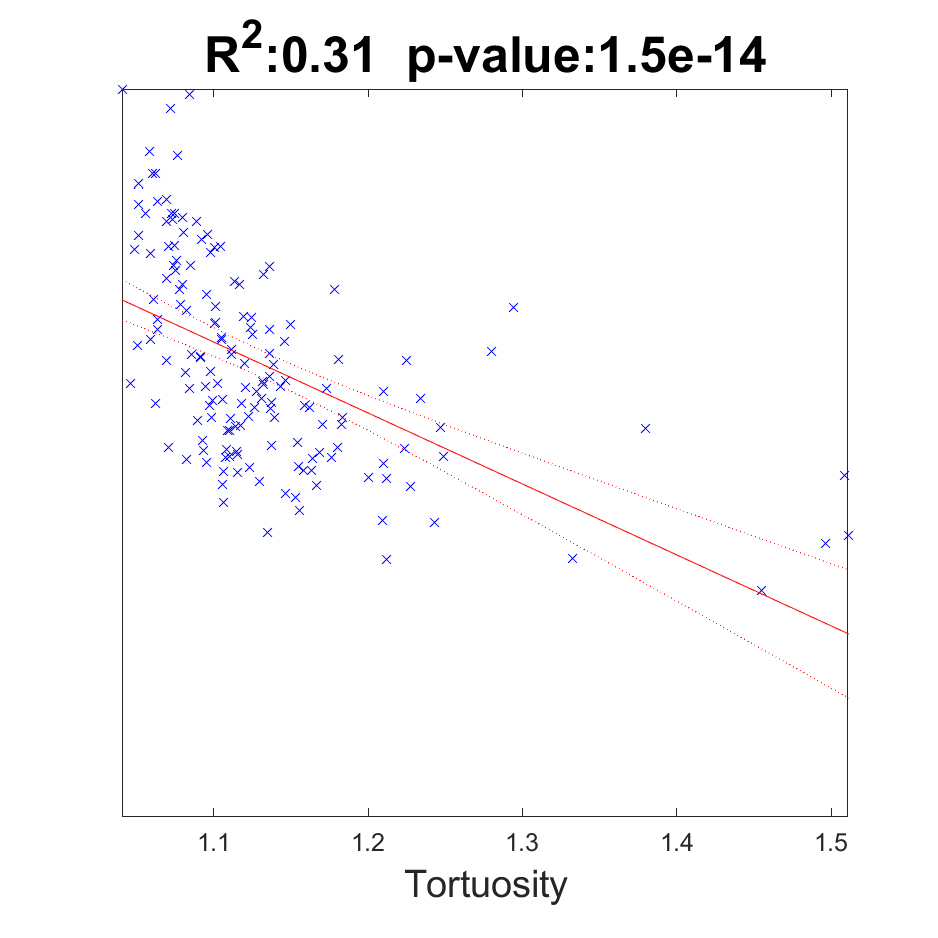}
         \caption{VOI4 Velocity-Tortuosity at 1.6m/s}
        \label{fig:my_label}
    \end{subfigure}
    \caption{Linear regression models created by considering average streamline velocity magnitude values from CFD simulations and architectural parameters obtained through IA (Image analysis). 
    (a) Strong correlation between velocity and Re for VOI(3) at low inlet velocity. (b) Good correlation between velocity and channel diameter for VOI(3) at low inlet velocity. (c) Low correlation between Velocity and tortuosity for VOI(3) at low inlet velocity. (d) Strong correlation between velocity and Re for VOI(3) at high inlet velocity. (e) Low correlation between velocity and channel diameter for VOI(3) at high inlet velocity. (f) Good correlation between Velocity and tortuosity for VOI(3) at high inlet velocity. (g) Strong correlation between velocity and Re for VOI(4) at low inlet velocity. (h) Good correlation between velocity and channel diameter for VOI(4) at low inlet velocity. (i) Low correlation between Velocity and tortuosity for VOI(4) at low inlet velocity. (j) Strong correlation between velocity and Re for VOI(4) at high inlet velocity. (k) Low correlation between velocity and channel diameter for VOI(4) at high inlet velocity. (l) Medium correlation between Velocity and tortuosity for VOI(4) at high inlet velocity.}
    \label{Figure: VOI Regression Data}
\end{figure}

To obtain an understanding of the structure-parameter relationship of the meniscus, statistical analysis was used help to explain fluid behaviour as a result of structural parameters.
Regression modelling has been used to find correlations between fluid velocity and a number of architectural parameters such as channel diameters and tortuosity. Fig. \ref{Figure: VOI Regression Data} show the results of the linear regression analysis for both VOI(3) and VOI(4):

\begin{itemize}

    \item At low speeds, channel diameter was the most influential parameter of stream velocity. Possibly due to the fact that larger channel experience less fluid friction around the centre of the channel.
    \item At high speeds, tortuosity becomes the driving factor of fluid velocity. This could be due to the fact that more energy is required to overcome these topological resistances. At lower speeds, this parameter is notably less impactful.
    \item Re number presents a consistently high correlation with fluid velocity, which is to be expected as it is a function of velocity. However, this relationship is seen to break down slightly at higher velocities. What causes this is still not fully known.
    \item Connectivity, while having been assessed, has not been presented as it presented very little correlation with fluid velocity, with $R^2$ ranging between 0-0.18. Indicating that inter-channel connectivity is not critical.
    
\end{itemize}

\subsection{Averaged architectural and flow parameters}          

In this section, results on the averaged architectural and flow parameters obtained from the analysis of VOI(3) and VOI(4) are presented. To study the spatial evolution of flow parameters i.e. Re and permeability, along with architectural parameters, VOI(3) and VOI(4) have been divided into 7 and 9 sections respectively, as described in Fig. \ref{fig:PorosityvsL} d. For each of the equally spaced sections, average architectural parameters are reported in Table\ref{table:archi}. It has been noted that VOI(3) shows porosity varying between 38\% (section 7) and 50\% in sections 5 and 6. The mean channel diameter also varies between 56$\mu$m and 62$\mu$m. The tortuosity varies between 1.154 in sect 7 and 1.411 in sect 4.  In a Darcian flow regime with low Re, the permeability calculations show an average for VOI(3) of 27D which is reinforced by calculations done using PNM, The VOI(4) average is similar at about 29D. VOI(3) exhibits a drop in permeability by 26\% in section 7 with respect to the average, consistent with a reduction in porosity in the same location highlighted above (38\% porosity). VOI(4) shows more uniform values of architectural parameters and flow parameters throughout the sections. It can be noted from Fig.\ref{fig:PorosityvsL}e that the distribution of permeability for VOI(4) has a peak around 30 D which is close to the overall permeability of the sample. However, at some locations of the sample permeabilities are determined to be as high as 80D (although this is based upon small volumes calculated with PNM method when generating the results shown in Fig. \ref{fig:PorosityvsL} E). 
In addition, Table \ref{table:flow} shows the average Reynolds number in the sections for the three inlet velocities: 0.0001m/s (Re=0), 0.02025m/s (Re=1-2) and 1.64025m/s (Re around 150). VOI(3) exhibits a peak of Re=157 corresponding to the highest inlet velocity of 1.64025m/s in Section 7 and a min of Re=144 min in Section 2, while Re in VOI(4) varies between 145 and 166. 
\subsection{Fluid flow regimes inside the meniscus}

\begin{figure}[h!]
    \centering
    \includegraphics[width=\textwidth]{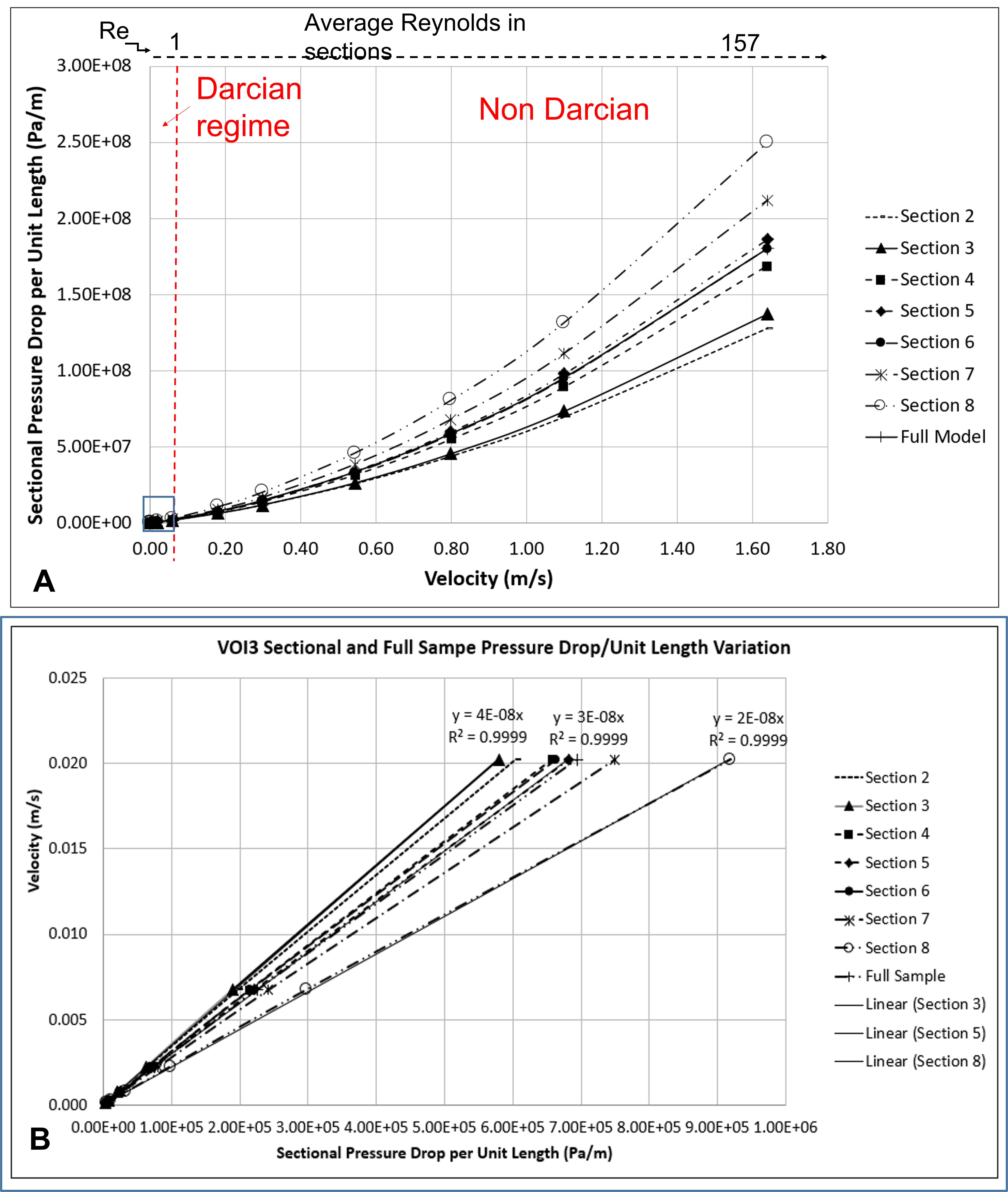}
    \caption{ A. Pressure drop per unit length from CFD analysis versus inlet velocity (from 0.1mm/s to 1.64m/s) calculated in the 7 sections inside VOI(3), showing the non-uniformity of the pressure drop in the domain.  We notice that the transition from Darcian to non-Darcian behaviour occurs at an inlet velocity of 20mm/s. B. Enlargement of the linear (Darcian) region of the graph used to calculate permeabilities values shown in table \ref{table:flow}.}
    \label{fig:OlgaNonDar}
\end{figure}

 Fig.\ref{fig:OlgaNonDar} shows the pressure drop/inlet velocity plot for each of the sections of VOI (3) similar to what is shown in Fig.\ref{fig:PorosityvsL}d which shows the transition between Darcian and Non darcian regime. It has been shown that the relationship is linear up to an inlet velocity of 0.02025m/s, corresponding to a mass flow rate of 7.678E-5$m^3/s$, and to a pressure drop per unit length varying from 0.6-0.9 MPa/m inside the sections of the domain. Section 7 shows the highest pressure drop per unit length (0.9 MPa/m) which corresponds to a pressure drop of 0.0036 MPa, Section 2 shows the lowest pressure drop per unit length  of 0.59 MPa/m  which corresponds to a pressure drop of 0.0023MPa. It can be noted that section 7 and section 2 present the same mean channel diameter (36$\mu$m), however section 7 presents the lowest porosity values 38\% (\ref{table:archi}). The average of the full sample in terms of pressure drops per unit length is 0.7 MPa/m corresponding to a pressure drop of 0.0021 MPa (0.02 bar). Beyond the inlet velocity of 0.02025m/s, we observe the transition to the non-Darcian or Forchheimer regime. Similar results are obtained for VOI(4).  

\begin{table}[]
\centering
\resizebox{\textwidth}{!}{%
\begin{tabular}{|cc|c|ccccccccc|}
\hline
\rowcolor[HTML]{9B9B9B} 
\multicolumn{1}{|c|}{\cellcolor[HTML]{9B9B9B}} &
  \cellcolor[HTML]{9B9B9B} &
  \cellcolor[HTML]{9B9B9B} &
  \multicolumn{9}{c|}{\cellcolor[HTML]{9B9B9B}\textbf{Sectional Re}} \\ \cline{4-12} 
\rowcolor[HTML]{9B9B9B} 
\multicolumn{1}{|c|}{\multirow{-2}{*}{\cellcolor[HTML]{9B9B9B}\textbf{\begin{tabular}[c]{@{}c@{}}Inlet \\ velocity $(m/s)$\end{tabular}}}} &
  \multirow{-2}{*}{\cellcolor[HTML]{9B9B9B}\textbf{\begin{tabular}[c]{@{}c@{}}Mass flow\\ rate $(m^2/s)$\end{tabular}}} &
  \multirow{-2}{*}{\cellcolor[HTML]{9B9B9B}\textbf{Sample}} &
  \multicolumn{1}{c|}{\cellcolor[HTML]{9B9B9B}\textbf{1}} &
  \multicolumn{1}{c|}{\cellcolor[HTML]{9B9B9B}\textbf{2}} &
  \multicolumn{1}{c|}{\cellcolor[HTML]{9B9B9B}\textbf{3}} &
  \multicolumn{1}{c|}{\cellcolor[HTML]{9B9B9B}\textbf{4}} &
  \multicolumn{1}{c|}{\cellcolor[HTML]{9B9B9B}\textbf{5}} &
  \multicolumn{1}{c|}{\cellcolor[HTML]{9B9B9B}\textbf{6}} &
  \multicolumn{1}{c|}{\cellcolor[HTML]{9B9B9B}\textbf{7}} &
  \multicolumn{1}{c|}{\cellcolor[HTML]{9B9B9B}\textbf{8}} &
  \textbf{9} \\ \hline
\rowcolor[HTML]{EFEFEF} 
\multicolumn{1}{|c|}{\cellcolor[HTML]{EFEFEF}} &
  \cellcolor[HTML]{EFEFEF} &
  VOI(3) &
  \multicolumn{1}{c|}{\cellcolor[HTML]{EFEFEF}7.12E-3} &
  \multicolumn{1}{c|}{\cellcolor[HTML]{EFEFEF}7.14E-4} &
  \multicolumn{1}{c|}{\cellcolor[HTML]{EFEFEF}7.72E-4} &
  \multicolumn{1}{c|}{\cellcolor[HTML]{EFEFEF}7.97E-3} &
  \multicolumn{1}{c|}{\cellcolor[HTML]{EFEFEF}7.71E-3} &
  \multicolumn{1}{c|}{\cellcolor[HTML]{EFEFEF}7.80E-3} &
  \multicolumn{1}{c|}{\cellcolor[HTML]{EFEFEF}8.66E-3} &
  \multicolumn{1}{c|}{\cellcolor[HTML]{EFEFEF}-} &
  - \\ \cline{3-12} 
\rowcolor[HTML]{EFEFEF} 
\multicolumn{1}{|c|}{\multirow{-2}{*}{\cellcolor[HTML]{EFEFEF}0.0001}} &
  \multirow{-2}{*}{\cellcolor[HTML]{EFEFEF}7.215E-8} &
  \cellcolor[HTML]{EFEFEF}VOI(4) &
  \multicolumn{1}{c|}{\cellcolor[HTML]{EFEFEF}6.94E-3} &
  \multicolumn{1}{c|}{\cellcolor[HTML]{EFEFEF}7.22E-3} &
  \multicolumn{1}{c|}{\cellcolor[HTML]{EFEFEF}7.36E-3} &
  \multicolumn{1}{c|}{\cellcolor[HTML]{EFEFEF}7.77E-3} &
  \multicolumn{1}{c|}{\cellcolor[HTML]{EFEFEF}7.87E-3} &
  \multicolumn{1}{c|}{\cellcolor[HTML]{EFEFEF}7.93E-3} &
  \multicolumn{1}{c|}{\cellcolor[HTML]{EFEFEF}8.21E-3} &
  \multicolumn{1}{c|}{\cellcolor[HTML]{EFEFEF}9.27E-3} &
  \cellcolor[HTML]{EFEFEF}8.59E-3 \\ \hline
\rowcolor[HTML]{C0C0C0} 
\multicolumn{1}{|c|}{\cellcolor[HTML]{C0C0C0}} &
  \cellcolor[HTML]{C0C0C0} &
  VOI(3) &
  \multicolumn{1}{c|}{\cellcolor[HTML]{C0C0C0}1.44} &
  \multicolumn{1}{c|}{\cellcolor[HTML]{C0C0C0}1.45} &
  \multicolumn{1}{c|}{\cellcolor[HTML]{C0C0C0}1.57} &
  \multicolumn{1}{c|}{\cellcolor[HTML]{C0C0C0}1.62} &
  \multicolumn{1}{c|}{\cellcolor[HTML]{C0C0C0}1.57} &
  \multicolumn{1}{c|}{\cellcolor[HTML]{C0C0C0}1.59} &
  \multicolumn{1}{c|}{\cellcolor[HTML]{C0C0C0}1.76} &
  \multicolumn{1}{c|}{\cellcolor[HTML]{C0C0C0}-} &
  - \\ \cline{3-12} 
\rowcolor[HTML]{C0C0C0} 
\multicolumn{1}{|c|}{\multirow{-2}{*}{\cellcolor[HTML]{C0C0C0}0.02}} &
  \multirow{-2}{*}{\cellcolor[HTML]{C0C0C0}7.678E-5} &
  VOI(4) &
  \multicolumn{1}{c|}{\cellcolor[HTML]{C0C0C0}1.43} &
  \multicolumn{1}{c|}{\cellcolor[HTML]{C0C0C0}1.47} &
  \multicolumn{1}{c|}{\cellcolor[HTML]{C0C0C0}1.51} &
  \multicolumn{1}{c|}{\cellcolor[HTML]{C0C0C0}1.58} &
  \multicolumn{1}{c|}{\cellcolor[HTML]{C0C0C0}1.57} &
  \multicolumn{1}{c|}{\cellcolor[HTML]{C0C0C0}1.62} &
  \multicolumn{1}{c|}{\cellcolor[HTML]{C0C0C0}1.64} &
  \multicolumn{1}{c|}{\cellcolor[HTML]{C0C0C0}1.68} &
  1.73 \\ \hline
\rowcolor[HTML]{EFEFEF} 
\multicolumn{1}{|c|}{\cellcolor[HTML]{EFEFEF}} &
  \cellcolor[HTML]{EFEFEF} &
  VOI(3) &
  \multicolumn{1}{c|}{\cellcolor[HTML]{EFEFEF}125} &
  \multicolumn{1}{c|}{\cellcolor[HTML]{EFEFEF}129} &
  \multicolumn{1}{c|}{\cellcolor[HTML]{EFEFEF}138} &
  \multicolumn{1}{c|}{\cellcolor[HTML]{EFEFEF}145} &
  \multicolumn{1}{c|}{\cellcolor[HTML]{EFEFEF}141} &
  \multicolumn{1}{c|}{\cellcolor[HTML]{EFEFEF}140} &
  \multicolumn{1}{c|}{\cellcolor[HTML]{EFEFEF}157} &
  \multicolumn{1}{c|}{\cellcolor[HTML]{EFEFEF}-} &
  - \\ \cline{3-12} 
\rowcolor[HTML]{EFEFEF} 
\multicolumn{1}{|c|}{\multirow{-2}{*}{\cellcolor[HTML]{EFEFEF}1.6}} &
  \multirow{-2}{*}{\cellcolor[HTML]{EFEFEF}1.184E-3} &
  VOI(4) &
  \multicolumn{1}{c|}{\cellcolor[HTML]{EFEFEF}133} &
  \multicolumn{1}{c|}{\cellcolor[HTML]{EFEFEF}134} &
  \multicolumn{1}{c|}{\cellcolor[HTML]{EFEFEF}138} &
  \multicolumn{1}{c|}{\cellcolor[HTML]{EFEFEF}146} &
  \multicolumn{1}{c|}{\cellcolor[HTML]{EFEFEF}146} &
  \multicolumn{1}{c|}{\cellcolor[HTML]{EFEFEF}151} &
  \multicolumn{1}{c|}{\cellcolor[HTML]{EFEFEF}152} &
  \multicolumn{1}{c|}{\cellcolor[HTML]{EFEFEF}157} &
  161 \\ \hline
\rowcolor[HTML]{9B9B9B} 
\multicolumn{2}{|c|}{\cellcolor[HTML]{9B9B9B}} &
  \cellcolor[HTML]{9B9B9B} &
  \multicolumn{9}{c|}{\cellcolor[HTML]{9B9B9B}\textbf{Sectional permeability (D)}} \\ \cline{4-12} 
\rowcolor[HTML]{9B9B9B} 
\multicolumn{2}{|c|}{\multirow{-2}{*}{\cellcolor[HTML]{9B9B9B}\textbf{\begin{tabular}[c]{@{}c@{}}Calculation\\ method\end{tabular}}}} &
  \multirow{-2}{*}{\cellcolor[HTML]{9B9B9B}\textbf{Sample}} &
  \multicolumn{1}{c|}{\cellcolor[HTML]{9B9B9B}\textbf{1}} &
  \multicolumn{1}{c|}{\cellcolor[HTML]{9B9B9B}\textbf{2}} &
  \multicolumn{1}{c|}{\cellcolor[HTML]{9B9B9B}\textbf{3}} &
  \multicolumn{1}{c|}{\cellcolor[HTML]{9B9B9B}\textbf{4}} &
  \multicolumn{1}{c|}{\cellcolor[HTML]{9B9B9B}\textbf{5}} &
  \multicolumn{1}{c|}{\cellcolor[HTML]{9B9B9B}\textbf{6}} &
  \multicolumn{1}{c|}{\cellcolor[HTML]{9B9B9B}\textbf{7}} &
  \multicolumn{1}{c|}{\cellcolor[HTML]{9B9B9B}\textbf{8}} &
  \textbf{9} \\ \hline
\rowcolor[HTML]{EFEFEF} 
\multicolumn{2}{|c|}{\cellcolor[HTML]{EFEFEF}} &
  VOI(3) &
  \multicolumn{1}{c|}{\cellcolor[HTML]{EFEFEF}31} &
  \multicolumn{1}{c|}{\cellcolor[HTML]{EFEFEF}32} &
  \multicolumn{1}{c|}{\cellcolor[HTML]{EFEFEF}28} &
  \multicolumn{1}{c|}{\cellcolor[HTML]{EFEFEF}27} &
  \multicolumn{1}{c|}{\cellcolor[HTML]{EFEFEF}28} &
  \multicolumn{1}{c|}{\cellcolor[HTML]{EFEFEF}25} &
  \multicolumn{1}{c|}{\cellcolor[HTML]{EFEFEF}20} &
  \multicolumn{1}{c|}{\cellcolor[HTML]{EFEFEF}-} &
  - \\ \cline{3-12} 
\rowcolor[HTML]{EFEFEF} 
\multicolumn{2}{|c|}{\multirow{-2}{*}{\cellcolor[HTML]{EFEFEF}\begin{tabular}[c]{@{}c@{}}Computational Fluid Dynamics\\ (STAR CCM+)\end{tabular}}} &
  VOI(4) &
  \multicolumn{1}{c|}{\cellcolor[HTML]{EFEFEF}24} &
  \multicolumn{1}{c|}{\cellcolor[HTML]{EFEFEF}25} &
  \multicolumn{1}{c|}{\cellcolor[HTML]{EFEFEF}30} &
  \multicolumn{1}{c|}{\cellcolor[HTML]{EFEFEF}32} &
  \multicolumn{1}{c|}{\cellcolor[HTML]{EFEFEF}32} &
  \multicolumn{1}{c|}{\cellcolor[HTML]{EFEFEF}30} &
  \multicolumn{1}{c|}{\cellcolor[HTML]{EFEFEF}28} &
  \multicolumn{1}{c|}{\cellcolor[HTML]{EFEFEF}28} &
  33 \\ \hline
\rowcolor[HTML]{C0C0C0} 
\multicolumn{2}{|c|}{\cellcolor[HTML]{C0C0C0}} &
  VOI(3) &
  \multicolumn{1}{c|}{\cellcolor[HTML]{C0C0C0}24} &
  \multicolumn{1}{c|}{\cellcolor[HTML]{C0C0C0}29} &
  \multicolumn{1}{c|}{\cellcolor[HTML]{C0C0C0}31} &
  \multicolumn{1}{c|}{\cellcolor[HTML]{C0C0C0}30} &
  \multicolumn{1}{c|}{\cellcolor[HTML]{C0C0C0}27} &
  \multicolumn{1}{c|}{\cellcolor[HTML]{C0C0C0}29} &
  \multicolumn{1}{c|}{\cellcolor[HTML]{C0C0C0}21} &
  \multicolumn{1}{c|}{\cellcolor[HTML]{C0C0C0}-} &
  - \\ \cline{3-12} 
\rowcolor[HTML]{C0C0C0} 
\multicolumn{2}{|c|}{\multirow{-2}{*}{\cellcolor[HTML]{C0C0C0}\begin{tabular}[c]{@{}c@{}}Pore Network\\ Modelling\end{tabular}}} &
  VOI(4) &
  \multicolumn{1}{c|}{\cellcolor[HTML]{C0C0C0}37} &
  \multicolumn{1}{c|}{\cellcolor[HTML]{C0C0C0}33} &
  \multicolumn{1}{c|}{\cellcolor[HTML]{C0C0C0}32} &
  \multicolumn{1}{c|}{\cellcolor[HTML]{C0C0C0}33} &
  \multicolumn{1}{c|}{\cellcolor[HTML]{C0C0C0}35} &
  \multicolumn{1}{c|}{\cellcolor[HTML]{C0C0C0}33} &
  \multicolumn{1}{c|}{\cellcolor[HTML]{C0C0C0}33} &
  \multicolumn{1}{c|}{\cellcolor[HTML]{C0C0C0}27} &
  27 \\ \hline
\end{tabular}%
}
\caption{Averaged flow parameters: Reynolds number and permeability. Reynolds number calculated for three inlet velocity 0.00001m/s, 0.02025m/s and 1.64m/s. We show that meniscus exhibits a Darcian regime up to 0.02025m/s. Variation of permeability is shown in all sections. Permeability values are calculated by using CFD and PNM for comparison.}
\label{table:flow}
\end{table}
  
\section{Discussion}

This work opens up a number of research streams in the area of soft tissues exhibiting high porosity and load-bearing capacity. There is currently no investigation into the fluid flow regimes inside the internal layer of the central body within the human meniscus, nor on the statistical correlation between the architectural and flow parameters. The reason for this is that it has only been in recent years, due to the advancement of high-resolution imaging techniques such as multiphoton microscopy and micro-CT scanning, that the new meniscal structure constructed by a network of collagen channels has been revealed. The Extracellular Matrix (ECM) found in meniscal tissue is comprised of 60\% - 70\% water content. Biphasic (solid and fluid phases) macroscopic models such as poroelasticity, help explain the role of the fluid (i.e. pore pressure) in the mechanical response of the tissue. One of the main assumptions of poroelastic models is the dependence of Darcy's law. Darcy's law is a phenomenological relationship between the rate of fluid flow through porous media and plays a principal role in applications within the studies of hydrology, soils and tissue mechanics \citep{philip1992flow}. The derivation of Darcy’s law from principle Navier–Stokes equations assumes a creeping, laminar, stationary (time-independent fluid flux) and incompressible flow of density ($\rho$). The invalidation of the stationary flow assumption within meniscal tissue has been described in \citep{barrera2021unified, bulle2021human, gunda2023fractional}. It has been shown that fluid flux in the meniscus is time-dependent for a constant applied pressure gradient. A fractional (time) poroelastic model, in which only two parameters are involved (the order of the time derivative and the 'anomalous' permeability) captures the flow within the meniscus well and enables the identification of associated parameters \citep{barrera2021unified, bulle2021human}. Recently, the fractional poroelastic model has been validated and results have shown that the meniscus can be considered as a transversely isotropic poroelastic material. This statement was made due to the fluid flow rate being approximately three times higher in the circumferential direction than in the radial and vertical directions in the body region of the meniscus \citep{gunda2023fractional}. One of the core limitations of fractional models is the missing link between fractional parameters and the architectural features of the material. This could be overcome by performing image-based pore-scale simulations of fluid flow in a deformable body, which are currently investigated in the hemodynamics of heart valve arterial walls \citep{SCHUSSNIG2022115489, wang2022fluid}. However, for soft load-bearing tissues such as cartilage and menisci, this has yet to be investigated. This work is the first step towards the image-based fluid-structure interaction simulations which can complete the cycle. This study has delved deeper into this subject by exploring the morphology of these collagen channels, along with statistical analysis of architectural parameters (porosity, channel size, connectivity and tortuosity) and flow parameters such as Re number and permeability in the body region of the meniscus.

\par

The validity of assuming creeping laminar flow and its limitation in large porosity media has been discussed in \citep{osti_5051394}. This assumption leads to a linear relationship between fluid flux and gradient of pressure, for example, $-\nabla(p)=\frac{\mu}{k} u$. When the creeping laminar flow ceases, the inertial force begins to have an effect and the velocity-squared term need to be considered, $-\nabla(p)=\frac{\mu}{k} u + \beta \rho u^2$. This is known as 'Forchheimer’s correction' which comes into effect when the flow is subjected to inertial forces at the microscopic scale. The coefficient $\beta$ depends on the microstructure of the porous
medium. In biomechanics, it is usually believed that
Darcy’s law is sufficient to model the flow inside tissues and there are several benchmark tests performed to estimate the elastic and flow properties of articular cartilage and other soft tissues, during which severe loading conditions may trigger the microscopic inertial effects that call for Forchheimer’s correction. In these circumstances, the estimation of permeability may be inaccurate, these inaccuracies could be a cause of confusion when comparing experimental results with those found in the literature. For instance, confined compression tests are performed by \citep{bulle2021human, gunda2023fractional} to evaluate aggregate modulus and permeability of the meniscal tissue, the applied load is 0.07 MPa which generates an outlet velocity of 5mm/s, values of anomalous permeability varying from $k=0.86-3.75\times10^-12 m^4/Ns^1-\beta$, with higher permeability in the circumferential direction. Simulations show that outlet velocity could increase up to 40mm/s for an applied load of 0.7MPa, based on a study by \citep{gunda2023fractional}. In \citep{Katsuragawa}, confined compression tests are performed using an applied pressure of 0.02MPa and permeability is reported to be $k=60\times10^-15 m^4/Ns$. In \citep{danso2015characterization}, indentation tests performed with cyclic loading between 10gr and 60gr, corresponding to an applied pressure of about 0.1 MPa, are performed in order to estimate permeability, which was calculated to be between $k=0.01-0.06\times10^-15 m^4/Ns$. In \citep{SEITZ201368}, confined compression tests are performed using an applied pressure of 0.1 MPa and permeability has values $k=3.37-4.6\times10^-{15} m^4/Ns$.

\par 

In this study, we propose the CFD-IA method to carry out pore-scale simulations to calculate permeability and study in detail the evolution of the flow inside the microstructure. Using this method, it is possible to identify local and microscopic inertial effects, for example, inlet conditions of velocity/pressure, for which channels outside of the laminar regime can be identified ($Re>1000$ shown in fig. \ref{Figure: VOI Velocity and Re Distribution}a). It was observed that increasing inlet velocity (comparable to simulating increased loading in the meniscus) resulted in a reduction of the $\kappa$ parameter by 0.4 on average. The $\kappa$ concentration parameter indicates the distribution of flow direction inside the sample, a higher $\kappa$ value representing a more constrained flow direction throughout the sample, and a decrease in $\kappa$ indicating that the flow paths deviate more frequently from the preferential direction. This result suggests that the meniscus acts as a hydraulic damping device by spreading the flow across more channels to limit the amount of pressure drop generated.  

\par

Image-based CFD simulations of Darcy's style experiments used to estimate permeability have been reported for the performances of carbon fibres and brain tissue in \citep{scandelli2022computation, yuan2022microstructurally}. In \citep{scandelli2022computation}, it is discussed that the creeping flow regime is valid for inlet velocities less than 0.1m/s and a Re $< 0.5$, within the study conducted by \citep{yuan2022microstructurally} an inlet velocity of 0.0025 m/s is used. Darcy's law is used to estimate permeability; it has been found that the transition between Darcian and non-Darcian regimes occurs at an inlet velocity of 0.02$m/s$, the corresponding average Re in the domain is between 1-2, corresponding to a pressure drop through the sample of 0.0021MPa (0.02bar). However, local Re inside channels can reach values up to 24 at 0.02$m/s$ and 1500 at 1.64$m/s$. Various studies in the literature report a number of average Re calculations for porous materials \citep{zeng2006criterion}, and the applicability of Darcy's law \citep{Sobieski2014DarcysAF}. It is reported that the upper limit for the applicability of Darcy’s law is between Re = 1 and Re = 10, others suggest that the transition from Darcian and non-Darcian regime occurs for $Re>10$ \citep{hassanizadeh1987high}. \\
This ambiguity raises a number of questions: a. Which model is appropriate in cases where a range of velocities under consideration exceeds the Darcian flow regime? b. Which Re number formula should be used? c. How should flow inhomogeneity be assessed inside the domain, such as local inertial effects and how these relate to local morphology d. How to tackle the characterisation of flow parameters in a non-periodic anisotropic media, where finding a characteristic or RVE does not have a trivial solution. These questions are further exacerbated by the discrepancies found between permeability values for meniscal tissue in the literature \citep{elmukashfi2021model}. Values can vary from 1-3 orders of magnitude, from $k=\times10^-15 m^4/Ns$ to $k=0.01-0.06\times10^{-12} m^4/Ns$, equivalent to $k=\times 10^{-6} - \times 10^{-3}$ Darcy. Our calculations based on CFD and PNM report a considerably higher value of permeability ranging from $k=20-32 D$. This could be due to a number of reasons which deserve to be investigated further. 

\par

One possible explanation surrounds the difference between running physical and numerical experiments. In the numerical and analytical experiments (CFD and PNM), a digital sample has been extracted from a 3D reconstruction of a freeze-dried sample. Moreover, the sample is extracted in the preferential orientation of the channels, where permeability is higher, as shown in \citep{bulle2021human, gunda2023fractional}. Additionally, physical experiments in which permeability is derived by curve fitting can carry a number of uncertainties:

\begin{itemize}
    \item A variety of experimental techniques are used, from nano-micro indentation to confined/unconfined tests, with inconsistent loading conditions, ranging from 0.01MPa to 0.1MPa. Furthering this lack of regulation, different sample geometries and volumes are evaluated, with discrepancies ranging from a few hundred of microns to 3-4$mm$ in length.
    
    \item Multiple models are used to fit experimental curves to estimations of permeability, all endorsing the validity of a Darcian behaviour, with the exceptions being in \citep{bulle2021human, gunda2023fractional}.
    
    \item There is no direct experimental measurement of fluid flow (velocity/pressure/Re) inside the sample in order to validate the use of Darcy's law for the range of applied loading.
\end{itemize}

This work aims to address the above points, it has been shown that the method presented here, CFD-IA, can indeed be used as a tool to understand transport behaviour in meniscal tissue and quantify relationships between fluid flow parameters with architectural features of the porous medium. This paper intends to inform and inspire methods for designing biomimetic structures aimed at meniscus substitute/repair. One first attempt of using a combined audio-visual approach to generate artificial architectures mimicking the native ones by using generative AI is given in \citep{10.3389/fmats.2023.1092647}.

\par

To conclude, this work based on CFD-IA aims to bridge the difficulties in observing and quantifying fluid flow velocity and pressure inside meniscal channels, while simultaneously relating fluid parameters such as local Re with architectural features of the tissue such as tortuosity and connectivity. We have demonstrated that there is a strong correlation between velocity with tortuosity of a fluid path at high inlet velocities and with channel diameter at low inlet velocity. Moreover, this gives an indication of the onset between Darcian and non-Darcian behaviour and statistical distributions of macroscopic Re number and architectural parameters.

\section{Material and Methods}
\subsection{Micro-computed scans of meniscal architecture}
A sample of dimension $9mm$ x $10mm$ x $12mm$ was extracted from the central body of the medial meniscus (fig.~\ref{fig:PorosityvsL}a) with a surgical knife and immediately freeze-dried in a Benchtop Freeze Dryer (FreeZone Triad Cascade, Labconco) following the procedure in \citep{bonomo2020procedure}. Micro–Computed Tomography ($\mu$CT) analyses were carried out with a $\mu$CT scanner (Skyscan 1272, Bruker Kontich, Belgium). The sample was placed in the instrument's rotating support and the images were acquired with a resolution of 6.25 microns, by setting a rotation step of 0.2. To obtain an appropriate contrast, an X-ray beam with an energy level of 40 kV, an intensity of 250 mA and a 0.25mm thick aluminium filter was used. The total scanning time was 2:36 hours/scan. The scanned dataset obtained after the acquisition step consisted of images (1640 x 2452 pixels) in 16-bit tiff format.
Once the sample scanning is complete (fig.~\ref{fig:PorosityvsL}b), the acquired images were reconstructed using software N-Recon (version 1.6.10.2), used for the elimination of artefacts and noise reduction in each slice. The program returned the reconstructed images of the horizontal cross sections (XY plane), referred to as "Raw images", in 8-bit depth with 256 grey levels. Four unique Volumes of Interest (VOIs) were extracted from varying regions of the scanned sample (fig.~\ref{fig:PorosityvsL}b) in the circumferential direction (the preferential orientation of channels). The four datasets were then processed and analyzed as described below.

\subsection{Statistical characterisation of meniscal architecture} 
\label{sect: method, characterisation}
To characterise the meniscal pore space, it is essential to first draw distinct boundaries between the void and the material (collagen) elements within the samples. This was accomplished using thresholding, the four samples mentioned throughout this paper were 'binarised' using a grayscale threshold value of 34 to draw these boundaries. Once the pore space is clearly defined, it is possible to characterise it. Further distinctions must be made throughout the pore space and assumptions must be made on how an individual pore is defined, as this could significantly impact the quality of results and conclusions drawn, therefore, the selection of an appropriate segmentation method is paramount. In accordance with this, this paper follows a well-established methodology for segmenting the pore spaces of complex and varied porous media, the 'Maximal-Ball Algorithm' (MBA) \citep{Dong2009}. This methodology follows a series of critical processes to define and label individual pore structures, abbreviated as follows:

\begin{enumerate}
    \item The pore space is first described as a collection of spheres, created using a distance transform.
    \item The number of spheres is minimised to only those that represent the most amount of information, also called 'Maximal-Inscribed Spheres' (MIS).
    \item These spheres are passed through a hierarchical sorting algorithm, which assigns unique numerical labels to the spheres, corresponding to their pore 'family'.
    \item The centroids of these spheres, now representing constituent elements of pores, are expanded into the remaining undefined regions. This study utilises an iterative expanding distance map method, but previous studies have incorporated similar region-growing algorithms \citep{Baychev2019, Yi2017}.
    \item With the entire pore space labelled, pore connectivity can be evaluated by inspecting the voxels around the perimeter of each pore to detect all 26 neighbouring voxels. If a voxel contains a neighbour that is associated with a different pore family, then these pores are considered to be neighbours, and hence, connected. 
\end{enumerate}

A segmented pore space allows for the extraction of statistical distributions of parameters within the pore space. Key morphological and architectural parameters such as pore size, connectivity and throat size/ length to name a few can be interpreted from this methodology. In addition to this, quantifying parameters along the flow direction can also be beneficial for explaining fluidic characteristics within discrete sections of the flow domain. Investigation of this kind requires a 2-Dimensional analysis, rather than the aforementioned 3-Dimensional method. However, the characterised pore space can be utilised to draw out 2D results. Using the labelled matrix created from the segmentation process, it is possible to evaluate the parameters at each individual image plane. This could alternatively be accomplished by sequentially segmenting and evaluating the image planes, however, this lacks the continuity of relating pores in previous planes to future planes, and it was found that between planes, slight differences in pore morphology could yield notably different sizes. Therefore, for continuity and consistency, it was decided to utilise the 3D analysis. Although they will not be extensively discussed in this report, the distributions of these parameters, in both 2D and 3D, can be found within the supplementary information in Fig. S1 \& S2.

\subsection{Governing systems of equations: compressible flow}

The fluid flux through the meniscus was modelled by discretising the pore space in each sample. The viscous flow through the pore space was captured using the Reynold's Averaged Navier Stokes equations (Equation \ref{eq: Navier Stokes Equations}) in the absence of body forces. Where $u$ is the fluid velocity, $p$ is the pressure, $\rho$ is the density and $\mu$ is the viscosity (assumed constant).

\begin{equation}
        \frac{\delta(\rho u)}{\delta t} + \nabla.(\rho uu) = - \nabla p + \nabla.(\mu \nabla u)
     \label{eq: Navier Stokes Equations}
\end{equation}

    
    %
    %

 
The boundary conditions imposed consisted of a constant velocity applied to the inlet and a fixed pressure of 101.325 kPa at the outlet to the fluid domain. A no-slip wall condition was imposed on the soft tissue walls, $\textbf{u} = 0$. Siemens Simcenter Star-CCM+ CFD code was used to solve the partial differential equations using a segregated flow solver with a SIMPLE (Semi-Implicit Method for Pressure Linked Equations) Pressure-Velocity coupling algorithm . 
    
    %
    %

\subsection{Numerical Darcy's experiments: Computational Fluid Dynamics Simulations}

Darcy's experimental tests \citep{darcy1856public} was reproduced numerically using Siemens Star-CCM+ CFD software to characterise the fluid flow through the porous meniscal tissue samples. The pressure drop across the porous media is measured at the top and at the bottom of the tank and used to determine the permeability.

 
\par 

Two Micro-CT data sets (VOI(3) and VOI(4)) were converted into solid 3D geometries using 3D Slicer (www.slicer.org). The Micro-CT image-based porous/solid domains of the meniscus were converted into volume meshes. Some isolated small non-contiguous fluid volumes were removed to avoid numerical instabilities.


\par

The meniscus samples were placed into a cylindrical flow domain, the domain inlet and outlet were two and six times the meniscus sample length respectively. This provided model stability and robustness with the imposed boundary conditions and ensured the flow was developed post-meniscus to avoid reversal at the outlet boundary. The cylindrical meniscus samples were inserted in the flow domain with the samples' length aligned with the flow direction of the generated flow domain. The fluid volume was discretised using a polyhedral cell type in order to capture the non-uniformity of the meniscus pore structure \citep{doi:10.1073/pnas.2105328118}. Volumetric mesh controls were used to reduce the overall cell count in the inlet and outlet regions of the model; comprising of polyhedral cells with a target size of 0.015mm, with volumetric controls set on the inlet and outlet region with a target cell size of 0.0375mm. 
Also included was a double prism layer on all wall surfaces. The following cell quality metrics were targeted for the discretised volume meshes: Skewness Angle $<$ 85 degrees; Face Validity $>$ 0.8; Volume Change $>1\times10^{-3}$.
The final CFD discretised geometries of VOI(3) and VOI(4) have $7.3\times10^6$ and $20.4\times10^6$ cells respectively. Fluid Density ($\rho$) is based on the IAPWS-IF97 (International Association for the Properties of Water and Steam, Industrial Formulation 1997) model which includes calculations for density and other thermodynamics properties based on the simulation pressure and temperature; the Water model was implemented in Star-CCM+. The dynamic viscosity of the fluid was 8.8871x10$^{-4}$ $Pa.s$ and, based on the boundary conditions, the density was 997$kg/m^3$.

\par 

To understand the flow structures of the meniscus samples and characterise the permeability, a range of inlet velocities were prescribed as the inlet to the fluid domain, varying from 1x$10^{-4}$ to 1.64025$m/s$.


The outlet pressure boundary was set to 101,325 Pa in order to identify the dependency on a range of possible meniscus compression loading-induced flow rates.

\subsection{Re number calculation in channels and Streamlines extraction} 
\label{sect: method, Re & streams}

To effectively combine the results of CFD analysis with the morphological and architectural results obtain through image analysis, it is first necessary to transfer the information from one platform to the other. To do this, the data of the flow simulations extracted from STAR-CCM+ were imported into MATLAB. STAR-CCM+ allows for the export of information at cell node locations, whether it be fluid velocity, pressure or any other parameters STAR provides. As exporting the information at every single node within the mesh would be an extensive and computationally expensive task, it was decided to export information from plane nodes at regular intervals of 18.75$\mu m$. This data can be relayed to the image data within MATLAB via the coordinates of the cell nodes. Therefore, it was imperative to ensure that the coordinates of the nodes suitably correspond to the same location within the image. This was confirmed in this work by probing the model in various within STAR and comparing it with the same coordinates within MATLAB. A visual comparison was used during this study, however, this could be refined in future to make use of spatial datums to more accurately relate relative positions to a set error. With the coordinate systems aligned, it was possible to import the CFD simulation data into MATLAB. Due to the fact that information was only extracted from STAR at discrete intervals, it was necessary to interpolate the values of the parameters at the voxels locations. MATLAB's built-in Scattered Interpolant function was used for this, a linear interpolation was employed for this study, it has been assumed that given the small intervals (approximately the equivalent of 3 image planes) that using a nonlinear method would not yield a significant difference.

\par

What is created from these interpolations is, in essence, a 3D image describing the flow through the domain. Combining this new information, in particular velocity, with the characterised pore space described in Section \ref{sect: method, characterisation}, it is possible to determine other fluid properties i.e. Reynolds number. This process can be explained by envisioning a single plane within the image stack. In a characterised pore plane, each pore channel is represented by a collection of labelled pixels. Given that these pixels have a known resolution, an associated area of the pore channel can be calculated. From the pore area (A), it is possible to stipulate an equivalent hydraulic diameter ($D_{hyd}$) by modelling the pore as a circle with an equivalent area, as shown in Equation \ref{eq: equiv_diam}.

\begin{equation}
    D_{hyd} = \sqrt{\frac{4A}{\pi}}
    \label{eq: equiv_diam}
\end{equation}

Each of the pixels used in the area calculation has an associated collection of pixels at the same location within images containing the data imported from the CFD simulations. Taking the data from these pixel locations, for example, velocity magnitude data, an average velocity of each pore channel can be derived by calculating the mean value of pixel velocities associated with each pore channel. With a hydraulic length and velocity, calculating the local Re number within each pore channel is simply described by Equation \ref{eq: re}, where $u$ = channel velocity ($m/s$) and $\nu$ = the kinematic viscosity of water at 25$^{\circ}C$ (8.93×10-7$m^2/s$). This simple process allows for the estimation of local Re in all pore cross-sections throughout the flow domain.

 \begin{equation}
    Re = \frac{u\times D_{hyd}}{\nu }
    \label{eq: re}
\end{equation}

Each channel cross-section can now be modelled as a constituent element, with an associated diameter, mean velocity and Re. Therefore, to calculate sectional values of fluid properties, the elements that fall within the specified Z-bounds are averaged to produce the results seen in Table \ref{table:archi}.

\par

Another tool within Star-CCM+ that can be exploited to great benefit through the use of image analysis is the incorporation of fluid streamlines. STAR provides the ability to select a start location (a point from which the streamline is released) and then generates an artificial trail representing the path of this point particle. This data provides insight into the dynamic trajectory of fluid inside the meniscal channels, from which we can infer conclusions about the overarching structure of the meniscus and relate what effect this structure has on fluid properties as demonstrated in Section \ref{sect: results}. The streamline tool, as it stands at the moment, is not designed to analyse the topological properties of media in detail but to instead extract and present simulation properties in more defined regions. Therefore, a bespoke algorithm was developed that interpreted the spatial coordinates of the streamlines, calculating topological properties such as tortuosity and channel orientation. The hydraulic streamline is comprised of a series of discrete points in 3D space, $S_{1} = (x_{1},y_{1},z_{1})... S_{n} = (x_{n},y_{n},z_{n})$. Therefore, it is possible to calculate the distance between each of these points as shown in Equation \ref{eq: point distance}, where $L_{i}$ is the Euclidean distance between two points. Hence, the straight line distance between the start and end of the stream is simply the Euclidean distance between the first and final point, and the sum of all these individual point distances is the total length of the stream, the ratio of which is the hydraulic tortuosity of the stream, as demonstrated in Equation \ref{eq: tort calc}.

\begin{equation}
    L_{i} = \sqrt{\sum (S_{i+1}-S_{i})^2} \equiv \sqrt{(x_{i+1}-x_{i})^2+(y_{i+1}-y_{i})^2+(z_{i+1}-z_{i})^2} 
    \label{eq: point distance}
\end{equation}

\begin{equation}
T_{Hyd} = \frac{\sum_{1}^{n-1} L}{\sqrt{(S_{n}-S_{1})^2}}
    \label{eq: tort calc}
\end{equation}

These streamlines also perform a dual function, working in conjunction with the previously described methodology to provide localised values of fluid properties. By tracing an individual streamline, it was possible to track the variation in properties, demonstrated by Figure \ref{Figure: VOI Velocity and Re Distribution}. Along with modelling the variability in fluid parameters, calculating the average of these parameters also provided valuable insight into the structure-function relationship of the meniscus. Modelling these average fluid property values against morphological, architectural and topological parameters as linear regressions made it possible to quantify these relationships, as demonstrated in Figure \ref{Figure: VOI Regression Data}. These linear regressions were computed using the 'fitlm' function from MATLAB's Statistics and Machine Learning Toolbox.

\subsection{Pore Network Model and permeability calculation}

Pore Network models are coarse-grid versions of the porous media's internal structure that can be used to characterise fluid transport with a considerably low computational cost compared to direct simulation techniques \citep{blunt2001flow,rabbani2019hybrid}. To find the absolute fluid permeability of the VOI(3) sample, a Watershed segmentation algorithm has been utilised to break down the pore space into a network of pores and throats \citep{rabbani2014automated} and solve the mass balance equation for each of the pores to find the fluid flowing pressure. In this pore network model, we have assumed that the laminar fluid flow is occurring at a very low velocity, due to the constant pressure boundaries at the input and output faces of the sample. The fluid conductivity of each of the throats is calculated using an image-based throat permeability model introduced in \citep{rabbani2019hybrid} which estimates the Lattice Boltzmann simulated permeability of the throat. By assuming a Darcy flow regime and writing the fluid flux conservation using the Hagen–Poiseuille equation, pore pressures can be found at the centre of each pore. This process is accomplished by forming a system of linear equations with only the pressure as unknown. Such a system of equations is solved using the biconjugate gradients method \citep{barrett1994templates}. On the basis of the calculated pore pressures, the flow rate of all throats will be determined and the absolute permeability of the whole sample can be obtained based on the Darcy equation. To find the tortuosity of the main flow pathways in the pore network model, a group of 2000 virtual particles is simulated to enter the inlet face and be carried by the fluid through the sample. The higher the flow rate of a throat, the higher the chance of a particle being carried through that throat. Then by analyzing the length of the dominant pathways, a flow-based tortuosity distribution can be calculated for the porous material (Fig. 1 E).

\bibliographystyle{elsarticle-num}
\bibliography{references}
\section{Acknowledgements}
O.B would like to acknowledge the European Union’s Horizon 2020 -EU.1.3.2. - Nurturing excellence by means of cross-border and cross-sector mobility under the Marie Skłodowska-Curie individual fellowship MSCA-IF-2017, MetaBioMec, Grant agreement ID: 796405. 

\begin{minipage}{0.5\textwidth}
    \includegraphics[width=\textwidth]{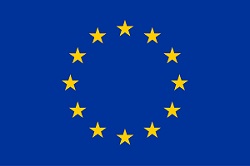}
\end{minipage}\hspace{10pt}
\begin{minipage}{0.5\textwidth}
    
    This paper only contains the author's views and the Research Executive Agency and the Commission are not responsible for any use that may be made of the information it contains.\\
\end{minipage}
\end{document}